\documentclass[final,1p,times]{elsarticle}
\usepackage{amssymb,amsmath,amsfonts}
\usepackage[toc,page]{appendix}
\usepackage{graphicx,psfrag,color}
\usepackage{dcolumn}
\usepackage{bm}
\usepackage{mathbbol}

\begin{document}

\begin{frontmatter}

\title{Almost perfect transmission of multipartite entanglement through 
disordered and noisy
spin chains}

\author{Rafael Vieira}
\ead{rafaelfis@df.ufscar.br}
\author{Gustavo Rigolin}
\ead{rigolin@ufscar.br}
\address{Departamento de F\'isica, Universidade Federal de
S\~ao Carlos, 13565-905, S\~ao Carlos, S\~ao Paulo, Brazil}


\begin{abstract}
We show how to efficiently send an $M$-partite entangled state along a spin chain of arbitrary size. Specifically, we show how an entangled $M$-partite W sate can be almost flawlessly transmitted from  one end (Alice) to the other end (Bob) of a spin-1/2 chain described by a slightly modified XX model. We achieve an almost perfect transmission without employing external magnetic fields or modulating the coupling constants among the spins of the chain, the two standard approaches used to achieve a good transmission efficiency. Moreover, the protocol here proposed can be used to transform an $M$-partite W state with Alice into an 
$\widetilde{M}$-partite one with Bob ($M\neq \widetilde{M}$).  
We also investigate the proposed protocol's response to 
several types of disorder and noise and show that it is quite robust to small 
deviations about the coupling constants of the optimal ordered and noiseless case.
\end{abstract}


\begin{keyword}
Quantum entanglement \sep Entanglement production \sep Quantum communication
\end{keyword}

\end{frontmatter}


\section{Introduction}

One of the main challenges to large-scale quantum computing and communication is the development of efficient quantum data transmission protocols \cite{ben00}. 
A spin chain is a promising platform leading to very efficient quantum communication protocols and which is particularly suited to connect the different components making a 
solid-state based quantum computer \cite{bos03}. Indeed, by using spin chains as the solution to quantum communication, we will be dealing with the same physical system to process and 
transmit quantum information in a solid-state based quantum computer.

We can roughly classify the several spin chain-based quantum communication protocols into
three groups. The first one is associated to those protocols whose main goal is the transmission of a single qubit along the chain \cite{bos03,chr04,alb04,chr05,nik04,sub04,woj05,kar05,shi05,har06,huo08,gua08,ban10,ban11,kur11,god12,apo12,lor13,apo13,kor14,shi15,pou15,zha16,che16,lof16,est17}. The second group contains those protocols aiming at the creation of entanglement between two specific qubits of the chain \cite{bos03,chr05,har06,ban10,ban11,lor13,apo13,est17b,li05,apo19}. The third and last group is related to those protocols specifically built to transmit multipartite states 
 from Alice to Bob (more than one qubit, for instance) \cite{har06,apo15,lor16,cir97,ple04,sem05,nic16,vie18,vie19,che19,apo19b}. 
See also Ref. \cite{mar16} for protocols that use the transmission of a quantum state as a way to implement a quantum logic operation. 

The main goal of this work is to generalize the bipartite quantum state transfer protocol presented in Refs. \cite{vie18,vie19} to the realm of multipartite states.  In particular, we present a quantum communication protocol targeted to transmit the genuine multipartite entangled $W$ state \cite{dur00}: $|W\rangle=(|100\cdots 0\rangle+|010 \cdots 0\rangle+...+
|000 \cdots 1\rangle)/\sqrt{M}$, where $M$ is the number of qubits forming the $W$ state. 
The state $W$ is an important quantum resource being suitable to, for instance, quantum secure communication \cite{joo02, wan07} and teleportation \cite{gor03}.

\begin{figure}[!ht] \begin{center}
\includegraphics[width=13cm]{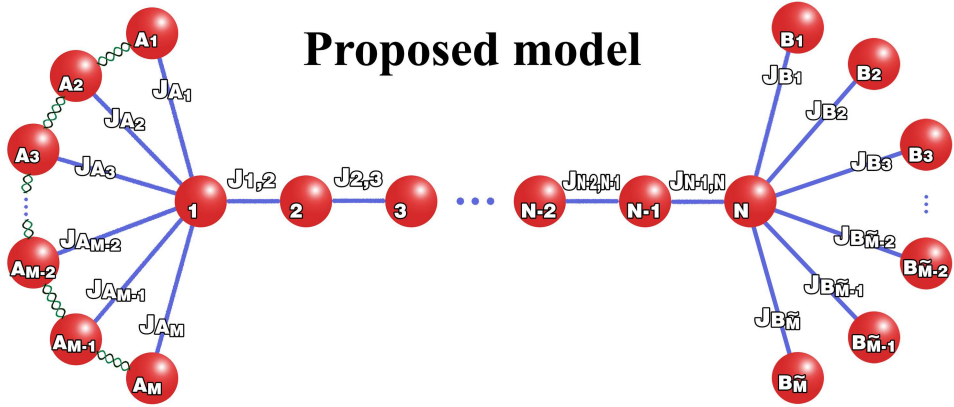}
\caption{\label{fig1}
Schematic representation of the proposed model. Initially all qubits $A_j$, $j=1,\cdots, M$,
with Alice are in the state $|W\rangle=(|100\cdots 0\rangle+|010 \cdots 0\rangle+...+|000 \cdots 1\rangle)/\sqrt{M}$ and all the other qubits are in the state $|0\rangle$. We will focus on unmodulated spin chains ($J_{i,i+1}=J_m$, $i=1,2,\ldots,N-1$) and our goal is to find the optimal constants $J_{\!_{A_i}}$ and $J_{\!_{B_i}}$ and the measurement time $t$ leading to the best entanglement transmission, i.e., we want the set of coupling constants and time $t$ for which Bob's 
$\tilde{M}$ qubits become an almost perfect $W$ state.}
\end{center} \end{figure}

The protocol here presented is scalable and flexible enough to transmit a $W$ state composed
of an arbitrary number of qubits. Irrespective of the number of qubits, we get an almost
perfect transmission from Alice to Bob for spin chain sizes ranging from hundreds to thousands of qubits. The protocol has a simple construction and operation as can be seen in Fig.~\ref{fig1}. All the qubits with Alice that constitute the $W$ state to be sent to Bob as
well as all the qubits with Bob that will receive the transmitted state 
do not interact with each other. They only interact with, respectively, 
the first and the last qubit of an $N$ spin one-dimensional chain. In order to differentiate
the qubits constituting the $W$ state from those of the spin chain, we call the former ``branches'', in analogy to the branches stemming from the trunk  
of a tree (the spin chain). We also employ the term ``quantum wire'' to designate the spin chain connecting Alice and Bob.

The present protocol has two features setting it apart from standard ways of implementing an  efficient quantum communication protocol via spin chains. First,
it is an unmodulated protocol, namely, we avoided any modulation in the coupling constants among the qubits along the chain \cite{chr04,nik04,kay05}. They are all equal and fixed
in time. Second, the interactions among the qubits of the spin chain are the sole responsible to drive the transmission of the quantum state from Alice to Bob. In our protocol, there is no need for external magnetic fields \cite{shi05,lor13,pem11,lor15}. Once a given $W$ state is prepared by Alice at 
the time $t=0$, we simply allow the dynamics of the system to deliver it to Bob at $t > 0$. In this scenario, we show that it is possible to adjust the coupling constants among the qubits such that an
almost perfect transmission is possible for the ordered and noiseless case. A different approach to transfer a genuine three-partite entangled state from Alice to Bob is given in Ref. \cite{apo19b}. In contrast to the present proposal, the authors of Ref. \cite{apo19b} 
send from Alice to Bob the $GHZ$ instead of the $W$ state.\footnote{The three-partite
$GHZ$ state can be written as $|GHZ\rangle=(|000\rangle+|111\rangle)/\sqrt{2}.$}

This paper is organized as follows. In Sec. \ref{tools} we present the mathematical 
concepts needed to a rigorous formulation of the present protocol. In Sec. \ref{branches} we show how to map the present model constituting of several branches to a pure linear chain. This allows us to explain why it works so well and how to properly set-up the optimal coupling constants among the spins, borrowing from the knowledge of the optimal couplings
associated with well-known strictly linear chain models (no branches). 
The concepts and quantities required to quantify the performance of the multipartite
entanglement transmission are shown in Sec. \ref{Fidelityandconcurrence}. 
Finally, in  Sec. \ref{robustness}, we study the robustness of the present model to noise and imperfections in its construction. This is done by introducing disorder to the optimal
set-up. Several types
and manners of introducing disorder are investigated \cite{vie18,vie19,vie13,vie14}, leading to
the conclusion that the present model is robust to small perturbations about the optimal values of the coupling constants. See also Refs.~\cite{chi05,fit05,bur05,pet10,zwi11,bru12,
nik13,kur14,ash15,pav16,ben03,ron16,lyr17,lyr17a,lyr17b} for more studies involving disorder
and noise.

\section{The proposed model} 
\label{tools}


The Hamiltonian of the present model is the isotropic XY model (XX model) with  
${M}$ bran\-ches interacting with qubit $1$ with Alice and $\tilde{M}$ branches coupled to qubit $N$ with Bob. We have a total of $N+M+\tilde{M}$ qubits and the Hamiltonian can
be written as follows,
\begin{equation}
H = H_A + H_N + H_B,
\label{ham0}
\end{equation}
where
\begin{eqnarray}
H_A  \hspace{-.2cm}&=&\hspace{-.2cm}  \sum_{p=1}^{{M}} J_{A_p} (\sigma_{A_p}^x\sigma_1^x+\sigma_{A_p}^y\sigma_1^y), \nonumber \\
H_N  \hspace{-.2cm}&=&\hspace{-.2cm}  \sum_{j=1}^{N-1}J_{j,j+1}(\sigma_j^x\sigma_{j+1}^x+\sigma_j^y\sigma_{j+1}^y), \nonumber \\
H_B  \hspace{-.2cm}&=&\hspace{-.2cm}  \sum_{q=1}^{{\tilde{M}}} J_{B_q} (\sigma_{N}^x\sigma_{B_q}^x\!+\!\sigma_{N}^y\sigma_{B_q}^y). \nonumber \\
\label{ham}
\end{eqnarray}
Note that $\sigma_i^\alpha\sigma_j^\alpha = \sigma_i^\alpha \otimes \sigma_j^\alpha$, with the superscript representing a particular Pauli matrix 
and the subscript fixing the qubit acted by it. 
We employ the following prescription, $\sigma^x|0\rangle=|1\rangle,\sigma^x|1\rangle=|0\rangle,\sigma^y|0\rangle=i|1\rangle,\sigma^y|1\rangle=-i|0\rangle, \sigma^z|0\rangle=|0\rangle,\sigma^z|1\rangle=-|1\rangle$, where 
$|0\rangle$ and $|1\rangle$ are the eigenstates of $\sigma^z$. In the 
up and down spin lingo,  $|\uparrow\rangle=|0\rangle$ and $|\downarrow\rangle = |1\rangle$. 
Also, if ${M}={\tilde{M}}=1$, ${J}_{A_1}=J_{B_1}=J$, and
$J_{j,j+1}=J$, $j=1,\ldots, N-1$, we obtain the Hamiltonian describing the standard 
(strictly linear) XX model
composed of $N+2$ qubits.

Since for this model the number of excitations is conserved \cite{vie18,vie19}, 
we can restrict ourselves to the one excitation subspace, where  
a general system of $N+M+\tilde{M}$ qubits 
is described by the superposition of $N+M+\tilde{M}$ states of one excitation.
Thus, following the notation of Refs.~\cite{vie18,vie19} and in accord with the 
nomenclature of Fig.~\ref{fig1}, at time $t$ the state describing our system is given by
\begin{equation}
|\Psi(t)\rangle = \sum_j c_{\!_j}(t)|1_j\rangle,
\label{psit}
\end{equation}
where $j={A_1},{A_2},\ldots, {A_{M}},1,2,3, \ldots, N,{B_1},{B_2},\ldots,{B_{\tilde{M}}}$ and
%
\begin{equation}
|1_j\rangle = \sigma_j^x|000\cdots\hspace{-.25cm} \underbrace{0}_\text{j-th qubit}\hspace{-.25cm} \cdots 000\rangle 
= |000\cdots\hspace{-.25cm} \underbrace{1}_\text{j-th qubit}\hspace{-.25cm} \cdots 000\rangle. 
\label{1j}
\end{equation}

At time $t=0$, Alice's qubits $A_1$, $A_2$, \ldots, $A_{M}$ are given by the generalized W state
\begin{equation}
|\psi(0)\rangle=|W\rangle=\frac{1}{\sqrt{{M}}}(|100\cdots 0\rangle+|010 \cdots 0\rangle+...+|000 \cdots 1\rangle)
\label{psi0}
\end{equation}
and the initial state of the system is 
\begin{equation}
|\Psi(0)\rangle=|\psi(0)\rangle \otimes |000\cdots0\rangle.
\label{psi1}
\end{equation}
Comparing Eqs.~(\ref{psi1}) and (\ref{psit})  we get 
\begin{eqnarray}
c_{\!_{A_1}}(0)&=&c_{\!_{A_2}}(0)=\cdots=c_{\!_{A_{M}}}(0)=1/\sqrt{{M}}, 
\label{initial1}\\
c_{\!_j}(0)&=&0, \hspace{.1cm} \mbox{for} \hspace{.1cm} j \neq A_1,A_2,...,A_{M}.
\label{initial2}
\end{eqnarray}
When $M={\tilde{M}}=1$, we have the standard model transmitting a single qubit state 
($c_{A_1}(0)=1$ and $c_{j}(0)=0 \hspace{.1cm} \mbox{for} \hspace{.1cm} j \neq A_1$).

Inserting $|\Psi(t)\rangle$, Eq.~(\ref{psit}), into the Schr\"odinger equation 
$$
i\hbar \frac{d}{dt}|\Psi(t)\rangle = H |\Psi(t)\rangle
$$ 
leads to 
\begin{equation}
i\hbar \frac{dc_{\!_k}(t)}{dt} = \sum_j c_{\!_j}(t) \langle 1_k|H|1_j\rangle
\label{ct}
\end{equation}
after taking the scalar product with the bra $\langle 1_k|$.

A straightforward but direct calculation gives
\begin{eqnarray}
\langle 1_k|H|1_j\rangle & = & 2 J_{\!_{A_1}}(\delta_{{A_1},k}\delta_{j,1}+\delta_{1,k}\delta_{j,{A_1}}) + 2 J_{\!_{A_2}}(\delta_{{A_2},k}\delta_{j,1}+\delta_{1,k}\delta_{j,{A_2}})  \nonumber \\
&& + \cdots + 2 J_{\!_{A_{M}}}(\delta_{{A_{M}},k}\delta_{j,1}+\delta_{1,k}\delta_{j,{A_{M}}})\nonumber \\
&& + 2 J_{\!_{B_1}}(\delta_{N,k}\delta_{j,{B_1}}+\delta_{{B_1},k}\delta_{j,N}) + 2 J_{\!_{B_2}}(\delta_{N,k}\delta_{j,{B_2}}+\delta_{{B_2},k}\delta_{j,N}) \nonumber \\
&& + \cdots + 2 J_{\!_{B_{\tilde{M}}}}(\delta_{N,k}\delta_{j,{B_{\tilde{M}}}}+\delta_{{B_{\tilde{M}}},k}\delta_{j,N}) \nonumber \\
&& + 2 \sum_{l=1}^{N-1} J_{l,l+1}(\delta_{l,k}\delta_{j,l+1}+\delta_{l+1,k}\delta_{j,l}),
\label{braket}
\end{eqnarray}
where $\delta_{j,k}$ is the Kronecker delta.

If we now insert Eq.~(\ref{braket}) into (\ref{ct}) we get
\begin{eqnarray}
\frac{dc_{\!_k}(t)}{dt} & = & -i\frac{2}{\hbar} \sum_{p=1}^{{M}} J_{\!_{A_p}}\delta_{1,k}c_{\!_{A_p}}(t) - i\frac{2}{\hbar}\left(J_{1,2}\delta_{2,k} + \sum_{p=1}^{{M}} J_{\!_{A_p}}\delta_{A_p,k}\right)c_{\!_1}(t) \nonumber \\
& & -i\frac{2}{\hbar}\sum_{j=2}^{N-1}(J_{j-1,j}\delta_{j-1,k}+J_{j,j+1}\delta_{j+1,k})c_{\!_j}(t) \nonumber \\
& & -i\frac{2}{\hbar}\left(J_{N-1,N}\delta_{N-1,k}+\sum_{q=1}^{{\tilde{M}}}J_{\!_{B_q}}\delta_{{B_q},k}\right)c_{\!_{N}}(t) -i\frac{2}{\hbar}\sum_{q=1}^{{\tilde{M}}} J_{B_q}\delta_{N,k}c_{\!_{B_q}}(t).
\label{difeq}
\end{eqnarray}

Equation (\ref{difeq}) is a system of $N+M+\tilde{M}$ linear first order differential equations with time independent  
coefficients, where
$k=A_1,A_2,\ldots,A_{M},1,2,3,\ldots,N,B_1,B_2,\ldots,B_{\tilde{M}}$. 

If we now define the column vector ($T$ means transposition)
\begin{equation}
\mathbf{c}(t) =
\left( c_{\!_{A_1}}(t), c_{\!_{A_2}}(t), \ldots, c_{\!_{A_{M}}}(t), c_{\!_1}(t), 
c_{\!_2}(t), \ldots, c_{\!_{N}}(t), c_{\!_{B_1}}(t), c_{\!_{B_2}}(t), \ldots, 
c_{\!_{B_{\tilde{M}}}}(t) \right)^T
\end{equation}
%
we can rewrite Eq.~(\ref{difeq}) as follows,
\begin{equation}
\frac{d\mathbf{c}(t)}{dt} = \mathbf{F} \,\, \mathbf{c}(t),
\label{mateq}
\end{equation}
where $\mathbf{F}$ is a matrix proportional to the single excitation sector of the Hamiltonian (see \ref{ap0}). 
The solution to Eq.~(\ref{mateq}) is $\mathbf{c}(t) = e^{\mathbf{F}\,\, t }\mathbf{c}(0)$, 
with $\mathbf{c}(0)$ being the column vector that represents the initial conditions in Eqs.~(\ref{initial1}) and (\ref{initial2}). It is important to note that $\mathbf{F}$ is an 
$(N+M+\tilde{M}) \times (N+M+\tilde{M})$ dimensional matrix and by using standard linear system numerical solvers we can compute the matrix exponential for chains of about $1000$ qubits without much computational effort. 

\section{Mapping to a strictly linear chain} \label{branches}

The first key observation we highlight is the fact that the Hamiltonian (\ref{ham0}), 
similar to the XX model, is such that it conserves the number of excitations during
the time evolution of the system \cite{vie18,vie19}. 
This implies that the dynamics of the system is 
restricted to the subspace of one excitation since the initial state $W$ has exactly one
excitation. 

Moreover, the model we will be dealing with is such that  
\begin{equation}
J_{A_1}=J_{A_2}= \cdots = J_{A_{M}}=J_A \hspace{.25cm}\mbox{and}\hspace{.25cm} 
J_{B_1}=J_{B_2}= \cdots = J_{B_{\tilde{M}}}=J_B.
\label{sameC}
\end{equation}
This means that all the branches with Alice interact with the same coupling constant $J_A$
with qubit 1 and all the branches with Bob interact with the same coupling constant $J_B$
with qubit $N$. If we define the permutation operators $\mathcal{P}^A_{ij}$, where $i$ and
$j$ stand for any pair of branches with Alice, and $\mathcal{P}^B_{ij}$, with now $i$ and
$j$ denoting any pair of branches with Bob, it is not difficult to see using Eq.~(\ref{ham0}) that 
\begin{equation}
[H,\mathcal{P}^A_{ij}] =  [H,\mathcal{P}^B_{ij}] = 0,
\end{equation}
where the square brackets denote the commutator. Since the Hamiltonian commutes with those 
permutation operators, their average values are conserved along the time evolution. 
Quantitatively it implies that 
\begin{eqnarray}
\frac{d}{dt}\langle\Psi(t)| \mathcal{P}^A_{ij}|\Psi(t)\rangle = \frac{1}{i\hbar}
\langle\Psi(t)| [\mathcal{P}^A_{ij},H] | \Psi(t)\rangle = 0 &\Longrightarrow& 
\langle\Psi(t)| \mathcal{P}^A_{ij}|\Psi(t)\rangle = \mbox{constant}, \label{pija}\\
\frac{d}{dt}\langle\Psi(t)| \mathcal{P}^B_{ij}|\Psi(t)\rangle = \frac{1}{i\hbar}
\langle\Psi(t)| [\mathcal{P}^B_{ij},H] | \Psi(t)\rangle = 0 &\Longrightarrow& 
\langle\Psi(t)| \mathcal{P}^B_{ij}|\Psi(t)\rangle = \mbox{constant}. \label{pijb}
\end{eqnarray}

Now comes the second key observation. The state $W$, Eq.~(\ref{psi0}), and consequently the initial state $|\Psi(0)\rangle$, Eq.~(\ref{psi1}), is an eigenstate of $\mathcal{P}^A_{ij}$
and $\mathcal{P}^B_{ij}$ with eigenvalue one: $\mathcal{P}^A_{ij}|\Psi(0)\rangle=|\Psi(0)\rangle$ and $\mathcal{P}^B_{ij}|\Psi(0)\rangle = |\Psi(0)\rangle$. This fact combined with Eqs.~(\ref{pija}) and (\ref{pijb}) imply that for any $t$ we have
\begin{equation}
\mathcal{P}^A_{ij}|\Psi(t)\rangle = |\Psi(t)\rangle \hspace{.25cm}\mbox{and}\hspace{.25cm}
\mathcal{P}^B_{ij}|\Psi(t)\rangle = |\Psi(t)\rangle. 
\end{equation}
In other words, not only are we restricted to the subspace of one excitation but also to a 
subspace within the subspace of one excitation spanned by the eigenvectors of the permutation operators $\mathcal{P}^A_{ij}$ and $\mathcal{P}^B_{ij}$ possessing eigenvalue one. On the other hand, for a system of $M$ qubits and at most one excitation, there are only two  eigenvectors of $\mathcal{P}_{ij}$ with eigenvalue one, namely, $|\hat{0}\rangle$ and the $W$ state itself, 
\begin{eqnarray}
|\hat{0}\rangle &=& |00\cdots 0\rangle = |0\rangle^{\otimes M}, \\
|\hat{1}\rangle &=& |W\rangle = \frac{1}{\sqrt{M}}\sum_{j=1}^{M}\sigma^x_j|\hat{0}\rangle.
\end{eqnarray}

Therefore, defining the following vectors,
\begin{eqnarray}
|\hat{0}_A\rangle = |0\rangle^{\otimes M}, & & |\hat{1}_A\rangle = \frac{1}{\sqrt{M}}\sum_{j=1}^{M}\sigma^x_j|\hat{0}_A\rangle, \label{hat0A}\\
|\hat{0}_B\rangle = |0\rangle^{\otimes \tilde{M}}, & & 
|\hat{1}_B\rangle = \frac{1}{\sqrt{\tilde{M}}}\sum_{j=1}^{\tilde{M}}\sigma^x_j
|\hat{0}_B\rangle, \label{hat1B}\\
|\hat{0}_N\rangle = |0\rangle^{\otimes N}, & & 
|\hat{1}_j\rangle = \sigma^x_j|\hat{0}_N\rangle \label{hat1j}, 
\end{eqnarray}
and assuming that Alice's state at $t=0$ is 
the $M$-partite $W$-state, the state describing the whole system at any $t$
is given by
\begin{equation}
|\Psi(t)\rangle = C_{100}(t)|\hat{1}_A\rangle|\hat{0}_N\rangle|\hat{0}_B\rangle
+\sum_{j=1}^NC_{0j0}(t)|\hat{0}_A\rangle|\hat{1}_j\rangle|\hat{0}_B\rangle + C_{001}(t)
|\hat{0}_A\rangle|\hat{0}_N\rangle|\hat{1}_B\rangle. \label{C}
\end{equation}
If we compare with Eq.~(\ref{psit}) we get 
\begin{eqnarray}
c_{j}(t)=C_{100}(t)/\sqrt{M} & \mbox{for}& j=A_1,\ldots,A_M, \label{c1}\\
c_j(t)=C_{0j0}(t) & \mbox{for}& j=1,\ldots,N, \label{c2}\\
c_{j}(t)=C_{001}(t)/\sqrt{\tilde{M}} & \mbox{for}& j=B_1
\ldots,B_{\tilde{M}}. \label{c3}
\end{eqnarray}
Naturally, at $t=0$ we have $C_{100}(0)=1$, with the other coefficients being zero, and for
a perfect transmission at $t=t_{\!_{MAX}}$ we will have $C_{001}(t_{\!_{MAX}})=1$ and the other 
coefficients zero.

Furthermore, using Eqs.~(\ref{hat0A}) and (\ref{hat1B}) we can define the operators
\begin{eqnarray}
\sigma_A^x = |\hat{0}_A\rangle \langle\hat{1}_A| + |\hat{1}_A\rangle \langle\hat{0}_A|, 
& \sigma_A^y = -i|\hat{0}_A\rangle \langle\hat{1}_A| + i|\hat{1}_A\rangle \langle\hat{0}_A|, 
& \sigma_A^z = |\hat{0}_A\rangle \langle\hat{0}_A| - |\hat{1}_A\rangle \langle\hat{1}_A|, 
\label{PA}\\
\sigma_B^x = |\hat{0}_B\rangle \langle\hat{1}_B| + |\hat{1}_B\rangle \langle\hat{0}_B|, 
& \sigma_B^y = -i|\hat{0}_B\rangle \langle\hat{1}_B| + i|\hat{1}_B\rangle \langle\hat{0}_B|, 
& \sigma_B^z = |\hat{0}_B\rangle \langle\hat{0}_B| - |\hat{1}_B\rangle \langle\hat{1}_B|, 
\label{PB}
\end{eqnarray}
which are nothing but the Pauli matrices expressed in the basis $\{|\hat{0}_A\rangle,|\hat{1}_A\rangle\}$ and $\{|\hat{0}_B\rangle,|\hat{1}_B\rangle\}$, respectively.
Note that the above Pauli matrices can also be written as
\begin{eqnarray}
\sigma_A^x = \frac{1}{\sqrt{M}}\sum_{p=1}^M\sigma^x_{A_p}, 
\hspace{.25cm} \sigma_A^y = \frac{1}{\sqrt{M}}\sum_{p=1}^M\sigma^y_{A_p},
\hspace{.25cm} \sigma_A^z = \frac{1}{\sqrt{M}}\sum_{p=1}^M\sigma^z_{A_p}, \label{PA2}\\
\sigma_B^x = \frac{1}{\sqrt{\tilde{M}}}\sum_{q=1}^{\tilde{M}} \sigma^x_{B_q}, 
\hspace{.25cm} \sigma_B^y = \frac{1}{\sqrt{\tilde{M}}}\sum_{q=1}^{\tilde{M}}\sigma^y_{B_q},
\hspace{.25cm} \sigma_B^z = \frac{1}{\sqrt{\tilde{M}}}\sum_{q=1}^{\tilde{M}}\sigma^z_{B_q}. \label{PB2}
\end{eqnarray}
To prove the equivalence between the two representations of these Pauli matrices, we simply compute their matrix elements in the basis $\{|\hat{0}_A\rangle,|\hat{1}_A\rangle\}$ and $\{|\hat{0}_B\rangle,|\hat{1}_B\rangle\}$ using the representation given by Eqs.~(\ref{PA})-(\ref{PB}) and by Eqs.~(\ref{PA2})-(\ref{PB2}).  
At the end we will see that both calculations lead to the same matrix elements, proving
therefore the equivalence between the two representations.

Using Eqs.~(\ref{PA2}) and (\ref{PB2}) and working with the basis defined by Eqs.~(\ref{hat0A}) and (\ref{hat1B}), it is not difficult to see that the relevant part of the Hamiltonian (\ref{ham0}) to the present problem, i.e., the permutationally symmetric sector with at most one excitation, is given by
\begin{equation}
H = H_A + H_N + H_B,
\label{hamN0}
\end{equation}
with
\begin{eqnarray}
H_A  \hspace{-.2cm}&=&\hspace{-.2cm}  J_A \sqrt{M} (\sigma_{A}^x\sigma_1^x+\sigma_{A}^y\sigma_1^y) = 
J_{A,1}(\sigma_{A}^x\sigma_1^x+\sigma_{A}^y\sigma_1^y), \label{hamAN} \\
H_N  \hspace{-.2cm}&=&\hspace{-.2cm}  \sum_{j=1}^{N-1}J_{j,j+1}(\sigma_j^x\sigma_{j+1}^x+\sigma_j^y\sigma_{j+1}^y), \label{hamNN} \\
H_B  \hspace{-.2cm}&=&\hspace{-.2cm}  J_B \sqrt{\tilde{M}} (\sigma_{N}^x\sigma_{B}^x\!+\!\sigma_{N}^y\sigma_{B}^y) = 
J_{N,B}(\sigma_{N}^x\sigma_{B}^x\!+\!\sigma_{N}^y\sigma_{B}^y). 
\label{hamBN}
\end{eqnarray}

Looking at Eqs.~(\ref{hamAN})-(\ref{hamBN}), we clearly see that we have an effective
Hamiltonian describing a strictly linear chain composed of $N+2$ qubits. 
The first qubit (qubit A) interacts with the second one (qubit $1$) with coupling 
constant $J_{A,1} =  \sqrt{M} J_A$. The last qubit (qubit $B$) interacts
with the one before the last (qubit $N$) with coupling 
constant $J_{N,B} =  \sqrt{\tilde{M}} J_B$. The other coupling constants between nearest
neighbor qubits are given by $J_{j,j+1}$, where $j=1,\ldots, N-1$.

Therefore, the models described by Eqs.~(\ref{hamN0}) and (\ref{ham0}) are connected 
in the following important sense. The optimal settings leading to the most efficient transmission of a single excitation along an $(N+2)$-qubit strictly linear chain
can be used to determine the optimal settings of the model 
here proposed for the transmission of an $M$-partite entangled $W$ state along an 
$N$-qubit chain. We can get the optimal settings for Hamiltonian (\ref{ham0}) from
(\ref{hamN0}) by the following prescription,
\begin{equation}
J_A \longrightarrow \frac{J_{A,1}}{\sqrt{M}}, \hspace{.25cm} J_B \longrightarrow 
\frac{J_{N,B}}{\sqrt{\tilde{M}}}, \hspace{.25cm} J_{j,j+1} \longrightarrow J_{j,j+1}.
\label{optJ}
\end{equation}
Note also that we do not need $M$ and $\tilde{M}$ to be equal in order to implement an optimal transmission. 
This means that we can use the present proposal not only to send
an entangled $M$-partite $W$ state from Alice to Bob but also to transform it into an 
$\tilde{M}$-partite $W$ state at Bob's location by simply working with the appropriate number of 
branches and the corresponding optimal coupling constants as given by Eq.~(\ref{optJ}).

It is worth mentioning that we can relabel the coefficients and
the kets appearing in Eq.~(\ref{C}) as follows,
\begin{eqnarray}
(C_{100},\ldots, C_{0j0},\ldots,C_{001}) &=& 
(\tilde{c}_1,\ldots, \tilde{c}_{j+1},\ldots,\tilde{c}_{N+2}), \label{l1}\\
(|\hat{1}_A\rangle|\hat{0}_N\rangle|\hat{0}_B\rangle, \ldots,
|\hat{0}_A\rangle|\hat{1}_j\rangle|\hat{0}_B\rangle,
\ldots,
|\hat{0}_A\rangle|\hat{0}_N\rangle|\hat{1}_B\rangle ) & = & (|1_1\rangle), \ldots, 
|1_{j+1}\rangle, \ldots, |1_{N+2}\rangle, \label{l2}
\end{eqnarray}
where $|1_j\rangle$ is given by Eq.~(\ref{1j}) with $j=1,\ldots, N+2$. With this notation
Eq.~(\ref{C}) becomes
\begin{equation}
|\Psi(t)\rangle = \sum_{j=1}^{N+2} \tilde{c}_j(t) |1_j\rangle.
\label{psitt}
\end{equation}
Equation (\ref{psitt}) is what one would expect for a general state describing 
a strictly linear chain of $N+2$ qubits restricted to the single excitation sector.

Before we move on, it is important to stress the following point. 
The above map connecting the present model to a strictly linear chain  
only works, as we have already shown, if the two conditions below are satisfied.
First, the initial state must be a single excitation eigenvector of the 
permutation operators $\mathcal{P}^A_{ij}$ and $\mathcal{P}^B_{ij}$. This implies that
Alice's initial state must be a $W$ state.\footnote{We believe it is possible to establish a mapping for
states with more than one excitation too, such as the two excitation Dicke state of Ref.~\cite{kie07},
provided those states satisfy the same symmetries of the $W$ state as given in the text.
However, as we increase the number of excitations, the computational complexity to simulate the 
time evolution of the system also increases.}
Second, the coupling constants of all Alice's branches to qubit $1$ must be equal as well as the coupling constants of Bob's branches with
qubit $N$.  Therefore, in order to study how disorder and noise affect  
the present model, we must necessarily work with the original Hamiltonian (\ref{ham0}).
Indeed, in order to assess how disorder and noise acting \textit{independently} at each branch affects the transmission of the $W$ state, and there are $M$ + $\tilde{M}$ coupling
constants that might change independently due to disorder and noise, we have to work with
Hamiltonian (\ref{ham0}). Working with the strictly linear system is equivalent to letting the $M$ coupling constants at Alice's and the $\tilde{M}$ coupling constants with
Bob be affected by disorder in the same way since we only have $J_A$ and $J_B$ instead of
the $M+\tilde{M}$ couplings of the original model. 

\section{Quantifying the efficiency of the transmission}
\label{Fidelityandconcurrence}

\subsection{Fidelity}
The fidelity quantifies how close or similar the state received by Bob is to that sent by Alice. Let us assume we are interested in assessing the similarity of 
the state with Bob at the time $t$ to the one prepared and sent by Alice at $t=0$. 
The relevant quantity needed to compute the fidelity is the reduced density matrix 
describing Bob's branches, namely, $\rho_{B}(t)$. This is computed by tracing out 
all but Bob's branches from $\rho(t)$, the density matrix describing the total system,
\begin{equation}
\rho_{B}(t) = \mbox{Tr}_{\overline{B_1B_2\cdots B_{\tilde{M}}}} [\rho(t)].
\label{rhoBtraco}
\end{equation}
Here the bar over $B_1B_2\cdots B_{\tilde{M}}$ tells us that we are tracing out from 
$\rho(t)=|\Psi(t)\rangle\langle\Psi(t)|$ all qubits with the exception of those with Bob.
The state $|\Psi(t)\rangle$ is given by Eq.~(\ref{psit}) and 
inserting it into Eq.~(\ref{rhoBtraco}) we get after a little algebra,
\begin{eqnarray}
\hspace{-.15cm}\rho_{B}(t) \hspace{-.15cm}&=&
\left(\hspace{-.12cm}
\begin{array}{ccccc}
1-\sum_{j=1}^{\tilde{M}}|c_{B_j}(t)|^2 \hspace{-.125cm}&\hspace{-.125cm} 0 \hspace{-.125cm}&\hspace{-.125cm} 0 & \cdots & 0 \\
0 \hspace{-.125cm}&\hspace{-.125cm} |c_{B_1}(t)|^2 \hspace{-.125cm}&\hspace{-.125cm} c_{B_1}(t)c_{B_2}^*(t) & \cdots &\hspace{-.125cm} c_{B_1}(t)c_{B_{\tilde{M}}}^*(t) \\
0 \hspace{-.125cm}&\hspace{-.125cm} c_{B_2}(t)c_{B_1}^*(t) \hspace{-.125cm}&\hspace{-.125cm} |c_{B_2}(t)|^2 \hspace{-.125cm}& \cdots &\hspace{-.125cm} c_{B_2}(t)c_{B_{\tilde{M}}}^*(t)  \\
\vdots \hspace{-.125cm}&\hspace{-.125cm} \vdots \hspace{-.125cm}&\hspace{-.125cm} \vdots \hspace{-.125cm}& \ddots &\hspace{-.125cm} \vdots  \\
0 \hspace{-.125cm}&\hspace{-.125cm} c_{B_{\tilde{M}}}(t)c_{B_1}^*(t) \hspace{-.125cm}&\hspace{-.125cm} c_{B_{\tilde{M}}}(t)c_{B_2}^*(t) \hspace{-.125cm}& \cdots &\hspace{-.125cm} |c_{B_{\tilde{M}}}(t)|^2  
\end{array}
\hspace{-.12cm}
\right), 
\label{rhoNB}
\end{eqnarray}
where we used the normalization condition $\sum_j|c_{\!_j}(t)|^2=1$ to arrive at the first 
matrix element above. Note that $*$ represents complex conjugation and we are writing 
the matrix $\rho_B(t)$ in the basis 
$\{|\hat{0}_B\rangle,|\hat{1}_{B_{\!_1}}\rangle, \ldots, |\hat{1}_{B_{\!_{\tilde{M}}}}\rangle\}$, where $|\hat{1}_{B_{\!_j}}\rangle = \sigma_{B_{\!_j}}^x|\hat{0}_B\rangle$
(cf. Eqs.~(\ref{1j}) and (\ref{hat1B})-(\ref{hat1j})). 
Looking at Eq.~(\ref{rhoNB}),
we realize that the time dependence of the coefficients $c_{B_1}(t)$, $c_{B_2}(t)$,\ldots, 
$c_{B_{\tilde{M}}}(t)$ are all we need to fully characterize $\rho_{B}(t)$. 
Those coefficients are obtained solving Eq.~(\ref{mateq}).

%

We are now in a position to calculate the fidelity of Bob's state at time $t$ with respect to 
Alice's input state, the $W$ state,
%
%
%
\begin{equation}
F(t) = \langle W|\rho_{B}(t)|W\rangle= \frac{1}{M}\sum_{p=1}^{M}\sum_{q=1}^{M} c_{B_p}(t)c_{B_q}^*(t).
\label{fb}
\end{equation}
Note that in Eq.~(\ref{fb}) we are assuming, without loss of generality, that $M=\tilde{M}$.
If $M\neq \tilde{M}$ we compute $F(t)$ using a $W$ state matching the same number of qubits
with Bob. If the output state $\rho_{B}$ is equal to the input state $|W\rangle$ we have
$F=1$ and if they are orthogonal $F=0$. 

As proved in the previous section (see Eq.~(\ref{C}) and the discussion below it), 
the fact that we are sending a $W$ state along the chain implies that in the 
absence of noise and disorder $c_{B_j}(t)$ are all equal at any time. Thus, if we
define $c_{B_j}(t)=c(t)$, Eq.~(\ref{fb}) becomes
\begin{equation}
F(t) =M|c_{B_j}(t)|^2 = M |c(t)|^2.
\end{equation}
When studying the noisy and disordered case, however, we must rely on Eq.~(\ref{fb}) to
compute the fidelity. 

We can also express the fidelity in terms of the coefficients describing the effective 
strictly linear chain onto which we have mapped the present model. 
Using Eqs.~(\ref{C}), (\ref{c3}), and (\ref{l1}) we get
\begin{equation}
 F(t) = |\tilde{c}_{N+2}(t)|^2.
\label{fid1}
\end{equation}
It terms of the effective strictly linear chain, Eq.~(\ref{fid1}) is interpreted as the 
probability of seeing at time $t$ 
the single excitation at the site $N+2$, namely, at the last qubit of 
the $(N+2)$-spin chain (the qubit with Bob). Furthermore, as expected Eq.~(\ref{fid1}) is independent of the values of $M$ and $\tilde{M}$, respectively Alice's and Bob's number of branches 
in the original model. As such, the fidelity 
$F(t)$ computed via the effective model is equal to the fidelity of the original model 
for any values of $M$ and $\tilde{M}$ whenever the corresponding coupling constants are
given by Eq.~(\ref{optJ}).



On top of the condition that all branches interact with the same strength with the appropriate qubit of the chain (see Eq.~(\ref{sameC})), we also introduce the following two simplifications
to the proposed model. First, we assume that it is unmodulated \cite{woj05,vie18,vie19}, 
i.e., all couplings among the qubits of the chain are equal,
\begin{equation}
J_{j,j+1}=J_m \hspace{.25cm}\mbox{for}\hspace{.25cm} j=1,2,\ldots,N-1.
\label{cond1}
\end{equation}
Second, the interaction strength between the branches with the end qubits of the chain are
the same,
\begin{equation}
J_A = J_B = J.
\label{cond2}
\end{equation}
Note that if $J_A \neq J_B$ the transmission efficiency is reduced, as can be seen when we add
disorder to the system (see  Sec.~\ref{robustness}).

Since the proposed model and its map onto a strictly linear chain are equivalent when no disorder or noise is present, and in this section we do not want to deal with noise and disorder, from now on we fix our attention at the strictly linear 
model in order to find the optimal constants leading to a perfect transmission of a single excitation. We want to obtain the optimal settings leading to a time $t=t_{\!_{MAX}}$ such that
the fidelity, Eq.~(\ref{fid1}), equals one. 
Once the optimal settings are obtained for the linear model, 
we get the optimal ones for the proposed
model by simply applying the prescription given in Eq.~(\ref{optJ}).

For arbitrary values of $J$ and $J_m$, the search for the optimal settings is implemented numerically. However, for $J_m >> J$ we can solve for the optimal coupling constants
analytically. This is possible because in this scenario we can get very 
simple analytic expressions for the fidelity (see \ref{apA}):
\begin{eqnarray}
F(t) \approx \sin^2\left[2\left(\frac{J}{J_m}\right)\frac{J t}{\hbar}\right],
\hspace{.25cm}\mbox{for $N$ even}, \label{fide-fido0}\\
F(t) \approx \sin^4\left[\frac{2}{\sqrt{N+1}}\frac{J t}{\hbar}\right],
\hspace{.25cm}\mbox{for $N$ odd}. \label{fide-fido1}
\end{eqnarray}
To obtain the 
optimal couplings for a given $t$ 
we simply solve
$F(t)=1$ in Eqs.~(\ref{fide-fido0}) and (\ref{fide-fido1}).

It is instructive to compute at the same level of approximation 
the fidelity of the qubit with Alice with respect to the
single excitation state. Using the same techniques explained in \ref{apA} we get
$F_A(t) \approx \cos^2\left[2\left(\frac{J}{J_m}\right)\frac{J t}{\hbar}\right]$
for $N$ even and $F(t) \approx \cos^4\left[\frac{2}{\sqrt{N+1}}\frac{J t}{\hbar}\right]$
for $N$ odd. We see that Alice's fidelity is in quadrature with Bob's. As time
goes by, the excitation goes back and forth between Alice and Bob. The same cyclic behavior
is seen in the proposed model when Alice is sending $W$ states to Bob. 
 
\begin{figure}[!ht] \begin{center}
\includegraphics[width=13cm]{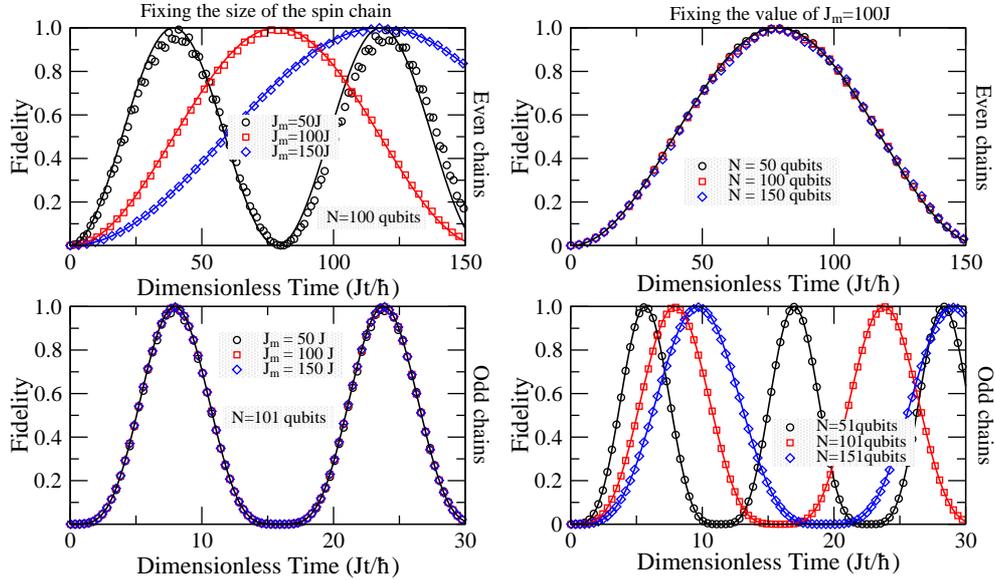}
\caption{\label{fig2}
All panels: Alice sends the genuinely entangled $M$-partite $W$ state to Bob, who 
receives it after a certain time $t_{\!_{MAX}}$ in the form of an almost perfect 
$\tilde{M}$-partite $W$ state ($F(t_{\!_{MAX}})\approx 1.0$),
with $M$ and $\tilde{M}$ arbitrary positive integers. The y-axes show the fidelity $F(t)$
between Bob's state and an $\tilde{M}$-partite $W$ state as a function of the 
dimensionless time $Jt/\hbar$. 
All curves and points were computed using the  
Hamiltonian (\ref{hamN0}), assuming unmodulated chains 
($J_{j,j+1}=J_m$ for $j=1,\ldots, N-1$) and $J_{A,1}=J_{N,B}=J$. 
See Secs. \ref{branches} 
and \ref{Fidelityandconcurrence} for details and why and how these results are
readily extended to arbitrary $M$ and $\tilde{M}$, i.e., valid for Hamiltonian 
(\ref{ham0}). The circle,
square, and diamond symbols are the exact values for the 
fidelity, computed numerically, while 
the solid curves were computed using the analytic formulas for the fidelity 
given by Eqs.~(\ref{fide-fido0}) and (\ref{fide-fido1}). In the upper-right and lower-left
panels all curves and symbols are indistinguishable.
Left panels: The chain's sizes connecting Alice and Bob
are fixed. Right panels: The value of $J_m/J$ is fixed. 
Upper panels: Chains with sizes given by an even $N$. 
Lower panels: Chains with sizes given by an odd $N$. 
Here and in the following figures all data are dimensionless.}
\end{center} 
\end{figure}

We have checked the accuracy of Eqs.~(\ref{fide-fido0}) and (\ref{fide-fido1}) 
by solving numerically for spin chains of sizes $N=50, 100$, and $150$ and for 
values of $J_m/J =50, 100$, and $150$. The agreement to the
analytic expressions is very impressive. Equations (\ref{fide-fido0}) and (\ref{fide-fido1}) fit almost perfectly onto the numerically computed 
points shown in Fig.~\ref{fig2}. 
Note that the greater the value
of $J_m/J$ the better those formulas 
are at describing the fidelity.

It is worth noticing that the prediction of Eq.~(\ref{fide-fido0}), telling us that 
for even $N$ the fidelity is independent of the size $N$ of the chain (upper-right panel 
of Fig.~\ref{fig2}), and
that of Eq.~(\ref{fide-fido1}), which says that for odd $N$ the fidelity is independent of $J_m$ (lower-left panel of Fig.~\ref{fig2}), hold
true already at the level of $J_m/J\approx 50$.

\subsection{Concurrence}

The fidelity is an excellent indicator of whether
or not the state received by Bob is close to the one sent by Alice.
This is particularly true for the ordered and noiseless proposed model, which 
can be mapped to a strictly linear chain transporting a single excitation. 
For the strictly linear chain we have seen that the
fidelity is nothing but the probability of Bob's
state being excited at a given time. Thus, the higher this probability the higher the
chances of the multipartite $W$ state of the original model 
being flawlessly transported to Bob.  

However, when disorder and noise are present the proposed model cannot be mapped to a 
strictly linear chain. In this scenario, it might
happen that no genuine multipartite entanglement reaches 
Bob in spite of reasonable values for the fidelity.
To make sure that genuine multipartite entanglement is reaching Bob, we need an 
entanglement measure suitably built to capture the specific type of 
genuine multipartite entanglement 
present in $W$-like states.

To accomplish this task, we note a very important property of an $\tilde{M}$-partite W state, namely, any pair of two qubits from a $W$ state are entangled \cite{dur00}. 
In other words, if we trace out $\tilde{M}-2$ qubits from the $W$ state, the remaining 
two qubits are entangled. Specifically, those two qubits has non-null concurrence,
an entanglement monotone devised to quantify bipartite entanglement \cite{woo98}. 

For any pair $i,j$ of qubits, obtained by tracing out the other $\tilde{M}-2$ qubits from 
$\rho_B$ (Eq.~(\ref{rhoNB})), we get for its concurrence \cite{woo98,vie18,vie19}
\begin{equation}
C_{i,j}=2|c_{B_{i}}(t)c_{B_j}(t)|. \label{conc1}
\end{equation}
Here $i\neq j$ and $i,j=1,\ldots,\tilde{M}$. Since for a $W$ state every pair of 
qubits has $C_{i,j}\neq 0$, the geometric mean of all $C_{i,j}$ 
is a natural quantity to test for and quantify 
genuine $W$-like entanglement. Calling $C_W$ this geometric mean we have,
\begin{equation}
C_W=\left(\prod_{i=1}^{\tilde{M}-1}\prod_{j>i}^{\tilde{M}} C_{i,j}\right)^{1/\mathcal{C}_{\tilde{M},2}}, \label{concurrence}
\end{equation}
where 
$$
\mathcal{C}_{\tilde{M},2} = \frac{\tilde{M}!}{(\tilde{M}-2)!2!}
$$ 
is the binomial coefficient. 

It is worth noticing that the use of the geometric mean is crucial to test for $W$-like entanglement. Indeed, if just a single pair of qubits has no entanglement at all, 
we immediately get $C_W=0$.
We can only have $C_W\neq 0$ if there is pairwise entanglement between all possible pairs
of qubits. This is the main feature of $W$-like genuine multipartite entanglement, a feature
neatly captured by the geometric mean of all $C_{i,j}$. For a $W$ state composed of 
$\tilde{M}$ qubits, we have for any pair $C_{i,j}=2/\tilde{M}$ and thus 
$C_W = C_{i,j} = 2/\tilde{M}$. For states that are not perfectly described by the 
$W$ state but that still have some $W$-like entanglement, we expect to have $0<C_W < 2/\tilde{M}$. If $C_W=0$, the state has no $W$-like entanglement at all.

We also define the following quantity,  
\begin{equation}
C_{i,j}^{MIN} = \min\{C_{1,2},C_{1,3},\ldots, C_{i,j},\ldots, C_{\tilde{M}-1,\tilde{M}}\},
\label{cij}
\end{equation}
which is the minimum value of pairwise entanglement (concurrence) available
among all pairs of qubits that can be formed from an 
$\tilde{M}$-partite $W$ state. This quantity will prove an important tool to 
roughly estimate the dispersion of the concurrence among all possible pairs of qubits.

\section{Robustness of the proposed model to disorder and noise} \label{robustness}

\subsection{Disorder}

Our goal now is to check the robustness of the present model to disorder. 
Specifically, we want to check how the entanglement transmission efficiency of the present
protocol is affected after we introduce random variations about the optimal values of the 
coupling constants that lead to an almost perfect transmission of genuine entanglement.

We can introduce disorder into our system in three different ways \cite{vie18,vie19,vie13,vie14}. 
First, we can randomly and independently change each one of the coupling constants about their optimal values before the transmission of the state. We do that only once 
and then let the system evolve until the time $t_{\!_{MAX}}$, the time when Bob would 
get an almost perfect replica of Alice's state had we employed the 
ordered model. At this time we
compute the fidelity of Bob's actual state with the $W$ state sent by Alice. 
This type of disorder is usually called \textit{static disorder}.
Second, we can also have \textit{dynamic disorder}. In this case we 
change all the coupling constants in the same way, let the system evolve
until the time $\tau$, change again all coupling constants in the same way 
at the time $2\tau$, and so forth. We keep repeating this procedure until we get to
$t_{\!_{MAX}}$. In this work, we choose $\tau = 10\%t_{\!_{MAX}}$, i.e., we change 
the coupling constants $10$ times before we get to $t_{\!_{MAX}}$. The third type 
of disorder here investigated, \textit{fluctuating disorder}, combines the features of
both static and dynamic disorders. In this scenario, the coupling constants are
independently and randomly changed about its optimal value at $t=0$ and along the 
time evolution. Those independent changes in the coupling constants during the time evolution are realized according to the prescription explained above for the dynamic disorder. 
See Ref.~\cite{vie19} for more details on how to numerically implement these types of disorder.

In this section we work with a spin chain of $N=100$ qubits and, 
without loss of generality, $M=\tilde{M}$, i.e., Alice and Bob have the same number of branches (see \ref{ap1} for the case where $N=1000$). 
As we did when studying the ordered model, we deal with unmodulated spin chains
and assume the same general settings for the coupling constants as explained in 
Sec.~\ref{Fidelityandconcurrence}.\footnote{We should mention that the map presented
in Sec.~\ref{branches} can also be implemented to modulated chains. Therefore, $W$ states
can also be efficiently transported using modulated chains to connect the branches 
in Fig.~\ref{fig1}. The optimal settings are the ones for the strictly 
linear modulated chain corrected by Eq.~(\ref{optJ}).}

For $N=100$ qubits, the optimal couplings in the ordered case and when we restrict 
ourselves to $\sqrt{M}J_m/J \leq 5.00$ is given by $\sqrt{M}J_m/J=2.03$. The optimal 
transmission fidelity in this case is $F=0.868$. For $\sqrt{M}J_m/J \leq 50.00$, the optimal value
is $\sqrt{M}J_m/J=49.39$, with $F=0.996$. 
In the numerical studies below, we set $J=1$ and
introduce disorder by changing $J_{A_p}=J_{B_q}=J/\!\sqrt{M}$
and $J_{j,j+1}=J_m$ about their optimal values.
Note that here $p,q = 1,\ldots, M$ and $j=1,\ldots, N-1$.  

We gauge how far we can go about the optimal settings by introducing the 
parameter $p$, which defines independent continuous uniform distributions centered in zero and ranging from $-p$ and $p$. For static disorder, 
each one of these continuous distributions is assigned to a given
coupling constant, changing 
$J_{i,j} \longrightarrow J_{i,j} (1 + \delta_{i,j})$, where $\delta_{i,j}$
is a number drawn from the corresponding uniform distribution. Here 
$J_{i,j}$ refers to $J_{A_p}$, $J_{B_q}$, or $J_{j,j+1}$. Note that we can think
of $p$ as representing the maximum percentage deviation of $J_{i,j}$ about its 
optimal value. For dynamical disorder, we only have one uniform distribution
such that $J_{i,j} \longrightarrow J_{i,j} (1 + \delta)$, where $\delta$
is a number drawn from this uniform distribution. At every interval of time 
$\tau$, we repeat the previous prescription, 
$J_{i,j} \longrightarrow J_{i,j} (1 + \delta)$, with $\delta$ being a different number
drawn from the same uniform distribution. Finally, for fluctuating disorder 
we have the same rule ascribed to the dynamical disorder with the following
modification. Now, 
each coupling constant has its own uniform distribution and, thus, at each
period $\tau$ we have $J_{i,j} \longrightarrow J_{i,j} (1 + \delta_{i,j})$, where 
$\delta_{i,j}$ is a number drawn from the uniform distribution corresponding 
to the coupling constant $J_{i,j}$.

\begin{figure}[!ht] \begin{center}
\includegraphics[width=13cm]{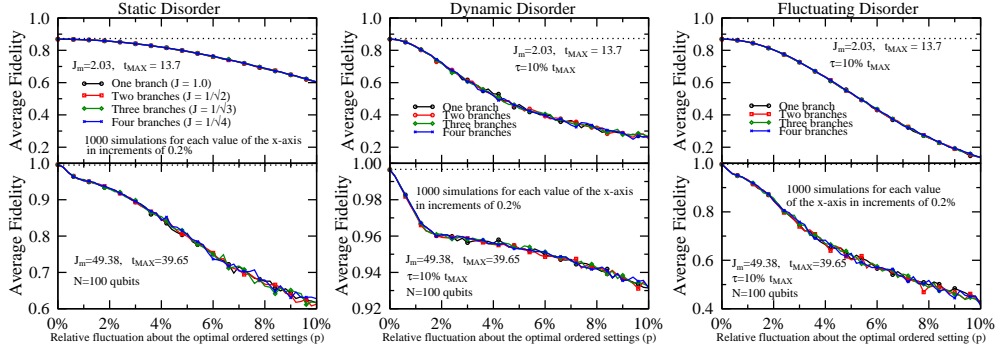}
\caption{\label{fig3}
Left: static disorder. 
Middle: dynamic disorder. Right: fluctuating disorder. We assume 
Alice and Bob have the same number of branches ($M=\tilde{M}$). The dotted lines show the fidelity for the ordered model. Upper panels: We realized 1000 simulations 
for each value of $p$ (percentage deviation about the optimal ordered case), 
from $p$=0.2\% 
to $p$ = 10\% in increments of 0.2\%, 
computing for each simulation the fidelity between Bob's 
state at $t_{\!_{MAX}} =13.7$ and
Alice's state at $t=0$.  The curves shown are the average fidelity after 1000
disorder simulations for each value of $p$. 
Note that $t_{\!_{MAX}}$ is the predicted time for 
optimal state transmission in the ordered system if $\sqrt{M}J_m/J = 2.03$, which is 
the best setting to transmit the $W$ state when $0\leq \sqrt{M}J_m/J\leq 5$.  
Lower panels: The
same as the upper panels but now $\sqrt{M}J_m/J=49.38$, which gives the best transmission for the ordered model when $0\leq \sqrt{M}J_m/J \leq 50$. In this case $t_{\!_{MAX}} = 39.65$. 
}
\end{center} 
\end{figure}

In Fig.~\ref{fig3} we show how the three different types of disorder affect the
transmission efficiency of the system. 
The first thing worth mentioning is that fluctuating disorder is the type of disorder
that affects most severely the system. The dynamical disorder, on the other hand, barely
affects the system for high values of $J_m$. For small values of $J_m$, however, 
the effect of dynamical disorder is almost as bad as that of fluctuating disorder. 
Moreover, for disorder strengths of the order of $p=1\%$, the system is not substantially
affected by any type of disorder, having in all cases a transmission 
fidelity of the order of $F=0.9$.  Also, the number of branches $M$ does not change
considerably the transmission efficiency of the protocol when all coupling constants
are acted by disorder.

\begin{figure}[!ht] \begin{center}
\includegraphics[width=13cm]{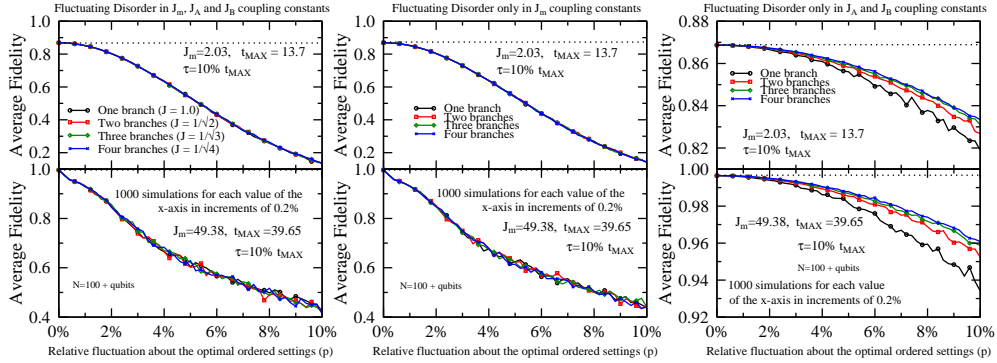}
\caption{\label{fig4}
The same as explained in Fig.~\ref{fig3} with the exception that now we deal only 
with fluctuating disorder. Fluctuating disorder acting simultaneously in 
$J_A$, $J_B$ and $J_m$ (left), just in $J_m$ (middle), and in $J_A$ and $J_B$ (right). 
}
\end{center} 
\end{figure}

Once we have determined that fluctuating disorder is the worst scenario, we now want 
to figure out which group of changing coupling constants are affecting most severely the system. The group represented by $J_{A_p}$, 
with $p = 1,\ldots, M$, we simply call $J_A$, the group of coupling constants given 
by $J_{B_p}$ we call $J_B$, and the coupling constants among the spins of the chain,
$J_{j,j+1}$, with $j=1,\ldots, N-1$, we call $J_m$.
As we can see looking at Fig.~\ref{fig4}, the fluctuations 
in $J_A$ and $J_B$ (right panels) are
considerably less important in determining the behavior of the disordered system than
those of $J_m$ (middle panels). Indeed, comparing the middle panels with the left ones
of Fig.~\ref{fig4}, we see that the decrease in the fidelity is dominated 
by the fluctuations of $J_m$. We can understand this feature by noting that the values of
$J_m$ are greater than the values of $J_A$ and $J_B$. Thus, the same percentage
fluctuation $p$ will lead to greater absolute changes for $J_m$, ultimately 
affecting more drastically the dynamics of the system than the small absolute changes of
$J_A$ and $J_B$.

Another interesting point that we can see looking at Fig.~\ref{fig4} is related to 
the case where only the branches with Alice and Bob are affected by disorder, i.e.,
only when the groups of coupling constants $J_A$ and $J_B$ are changed by disorder. 
Looking at the right panel of Fig.~\ref{fig4}, we see that the greater the number of 
branches the less susceptible to disorder is the system. This comes about because 
in the ordered model, the 
greater the number $M$ of branches the lower the interaction strength ($J_A=J_B=J/\sqrt{M}$)
between the branches and the endpoints of the chain  
that leads to an optimal transmission of the $W$ state.
Hence, for the same percentage fluctuation $p$, low values of $J_A$ and $J_B$ will lead
to small absolute changes in those coupling constants, affecting considerably less the 
dynamics of the system.

\begin{figure}[!ht] \begin{center}
\includegraphics[width=13cm]{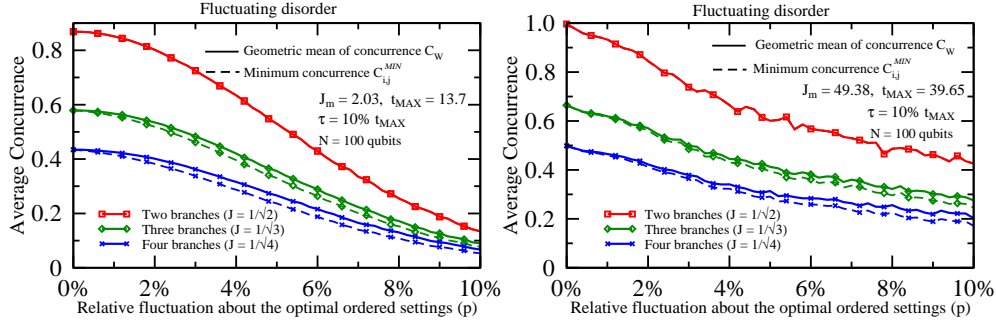}
\caption{\label{fig5}
Same as Figs. \ref{fig3} and \ref{fig4} but now we compute for each realization of 
disorder $C_W$  (see Eq.~(\ref{concurrence})), 
the geometric mean of the concurrence 
between all possible pairs of qubits (branches)
with Bob at the time $t_{\!_{MAX}}$. 
The solid curves are the average of $C_W$ 
after 1000 realizations of fluctuating disorder.
The smallest concurrence between all possible pairs of qubits with Bob, $C_{i,j}^{MIN}$, is given by the dashed lines. The square-red, diamond-green, and x-blue curves 
refer to spin chains with two, three, and four branches, respectively.
}
\end{center} 
\end{figure}

In order to be sure that genuine W-like multipartite entanglement is actually reaching 
Bob when disorder is present, we compute $C_W$ and $C^{MIN}_{i,j}$ in the worst disorder
scenario, namely, fluctuating disorder.
As defined in Sec. \ref{Fidelityandconcurrence}, the first quantity, 
Eq.~(\ref{concurrence}), is the geometric mean of all pairwise entanglement present in 
Bob's branches. The second quantity, Eq.~(\ref{cij}), picks the lowest pairwise entanglement present in 
a given pair of qubits with Bob. 
As discussed in Sec. \ref{Fidelityandconcurrence}, 
whenever $C_W\approx C^{MIN}_{i,j} \neq 0$, we are sure
that W-like entanglement is present.

Looking at Fig.~\ref{fig5}, we see that for small disorder ($p\approx 1\%$) the 
values of $C_W$ and $C^{MIN}_{i,j}$ are barely distinguishable. Moreover, for 
$\sqrt{M}J_m/J$ small (left panel), the values of $C_W$ and $C^{MIN}_{i,j}$
are almost the same as those predicted for the $W$ state. These features clearly illustrate
that genuine W-like multipartite entanglement is indeed reaching Bob for small disorder.
As we start increasing disorder, we note that the $C_W$ and $C^{MIN}_{i,j}$ decreases.
However, even as we increase the strength of disorder, we always have 
$C_W \approx C^{MIN}_{i,j}\neq 0$ all the way up to $p\approx 10\%$. This means that all pairs of qubits with Bob have almost the same level of pairwise entanglement. The dispersion 
in the values of concurrence is really low. Although for 
strong disorder we obviously do not have the predicted values of  
$C_W$ and $C^{MIN}_{i,j}$ for the pure $W$ state, we believe that since 
$C_W \approx C^{MIN}_{i,j}\neq 0$ we still have $W$-like entanglement present in 
Bob's qubits. 

\subsection{Noise}

We can introduce at least two types of noise in the proposed model which keep us in 
the one excitation subspace. Thus, all the numerical techniques employed so far 
can still be successfully used to investigate the robustness of the present model 
to these two particular types of 
noise \cite{vie18,vie19}. We can either introduce 
$\sigma_j^z\sigma_{j+1}^z$ interactions between the qubits of the chain as well as 
external magnetic fields acting on all spins along the $z$-axis.
Following the notation of Ref.~\cite{vie19}, these two types of noise are modeled
by adding the two terms below to Hamiltonian (\ref{ham0}),
\begin{eqnarray}
H_{zz} \hspace{-.2cm}&=&\hspace{-.2cm}  \sum_{p=1}^{{M}} \Delta_{A_p,1} \sigma_{A_p}^z\sigma_1^z + \sum_{j=1}^{N-1}\Delta_{j,j+1}\sigma_j^z\sigma_{j+1}^z + \sum_{q=1}^{{\tilde{M}}} \Delta_{N,B_q} \sigma_{N}^z\sigma_{B_q}^z\!,
\label{hamzz} \\
H_{z} \hspace{-.2cm}&=&\hspace{-.2cm}  \sum_{p=1}^{{M}} h_{A_p}\left(\mathbb{1}- \sigma_{A_p}^z\right) + \sum_{j=1}^{N}h_{j}\left(\mathbb{1}- \sigma_{j}^z\right) + \sum_{q=1}^{{\tilde{M}}} h_{B_q} \left(\mathbb{1}- \sigma_{B_q}^z\right).
\label{hamz}
\end{eqnarray}
Here $\Delta_{i,j}$ and $h_{j}$ represent, respectively, 
the coupling constant between nearest neighbor 
qubits affected by the $\sigma_j^z\sigma_{j+1}^z$ interaction and  
the strength of the coupling of the qubits with the external magnetic fields.

Similarly to the introduction of disorder to the proposed model, we can in the same
fashion deal with noise. We will have, therefore, static, dynamic, and fluctuating 
noise. The only difference is that at $t=0$ we have $\Delta_{i,j}=h_{j}=0$,
with $i,j$ assuming the appropriate values given in Eqs.~(\ref{hamzz}) and (\ref{hamz}).
Being more specific, for static noise we have $h_{j} \longrightarrow h_{j} + \delta_j$ and 
$\Delta_{i,j} \longrightarrow \Delta_{i,j}+ \delta_{i,j}$ only once at $t=0$. 
For dynamic noise, at every time $t$
that is an integer
multiple of the period $\tau$ we apply the prescription $h_{j} \longrightarrow h_{j} + \delta$ and 
$\Delta_{i,j}\longrightarrow \Delta_{i,j}+ \delta$, while for fluctuating noise at 
each multiple of the period $\tau$ 
we have $h_{j} \longrightarrow h_{j} + \delta_j$ and 
$\Delta_{i,j} \longrightarrow \Delta_{i,j}+ \delta_{i,j}$.
Note that $\delta_j$ and $\delta_{i,j}$ are random numbers drawn from 
a continuous uniform distribution ranging between -$p$ and $p$. Thus, when dealing with
noise, we can interpret $p$ as the maximal percentage fluctuation of 
$\Delta_{i,j}$ and $h_{j}$ from an interacting strength of unity value, 
i.e., $\Delta_{i,j}=h_j=1$.

\begin{figure}[!ht] \begin{center}
\includegraphics[width=8cm]{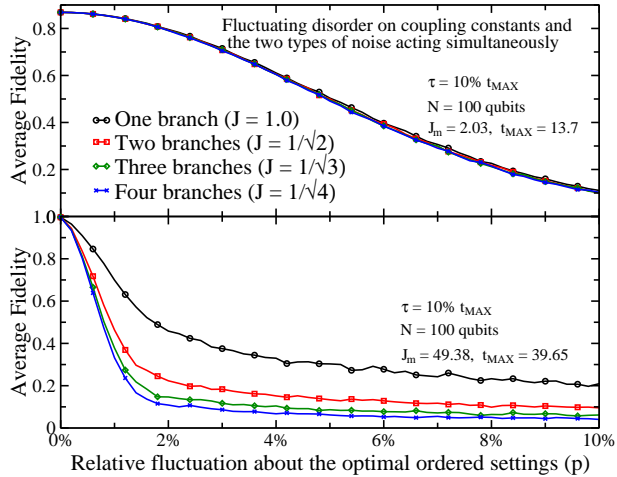}
\caption{\label{fig6b}
The same as Fig.~\ref{fig4} with the addition of two types of noise, namely,
the $\sigma_i^z\sigma_{j}^z$ interaction and external magnetic fields coupled 
to the spins of the system. The two types of noise act 
simultaneously in the same fashion as fluctuating disorder.  
}
\end{center} 
\end{figure}

In Fig.~\ref{fig6b} we present the worst case possible, i.e., fluctuating noise and 
fluctuating disorder on the coupling constants simultaneously present in the system.
Also, we work with the two types of noise acting at the same time on the system. 
Comparing Fig.~\ref{fig6b} with the left panel of Fig.~\ref{fig4},
we note that for small values of $J_m/J$ the effect of noise is negligible, barely 
affecting the fidelity of the transmitted state (efficiency). 
On the other hand, for high values
of $J_m/J$, the introduction of noise affects 
the transmission of the $W$ state in two important ways. 

First, the efficiency is drastically reduced in the presence of noise. At the $p=2\%$ level of noise and 
disorder, the fidelity $F$ of Bob's state is already of the order of $0.2$, while  
with only disorder in the coupling constants
we still have $F\approx 0.8$  (lower-left panel of Fig.~\ref{fig4}). 
Moreover, without noise we get for 
strong disorder ($p=10\%$) a fidelity of the order of $0.4$, which should be compared to
an almost null fidelity when we have both noise and disorder at $p\approx 10\%$.

Second, the introduction of noise for high values of $J_m/J$ ``breaks the degeneracy''
of the behavior of the fidelity as a function of the number of branches $M$ with Alice
and Bob. In the presence of noise, 
the greater $M$ the lower the fidelity 
for a given value of disorder and noise strength $p$ (lower panel of Fig.~\ref{fig6b}).
For $J_m/J\approx 50$, Bob's fidelity can be set approximately to $0.8$ if we 
only work with noise and disorder strengths not greater than $0.4\%$.

\section{Conclusion}
\label{conclusion}

In this work we extended the bipartite entanglement transmission protocol presented 
in Ref. \cite{vie18} to the domain of multipartite entanglement transmission. 
We showed how Alice can send almost flawlessly to Bob an $M$-partite
entangled state containing one single excitation without employing the standard and experimentally
demanding techniques to transmit quantum states along a spin chain, namely, a 
modulated spin chain or external magnetic fields to drive the state from Alice to Bob.
The fact that no external magnetic fields or modulation in the coupling constants 
among the spins are needed makes this protocol simpler to experimental implementations.

Specifically, 
in the present proposal Alice sends to Bob the genuinely multipartite entangled 
$W$ state: $|W\rangle=(|100\cdots 0\rangle+|010 \cdots 0\rangle+\cdots+|000 \cdots 1\rangle)/\sqrt{M}$. This state is encoded in $M$ qubits that do not interact 
among themselves after its preparation. Subsequently, at the time $t=0$ 
these qubits interact
individually and equally with one of the end points of a spin chain described by the XX model. The $N$ qubits of the spin chain as well the $M$ ones with Bob are prepared 
in the state $|0\rangle$ at $t=0$. 
By properly adjusting the interaction strength 
of Alice's and Bob's qubits with the end points of the spin chain
to the same predetermined value, we showed that the $W$ state is transmitted from
Alice to Bob solely due to the internal dynamics of the system. 
After a certain time $t>0$
the $M$ qubits with Bob become an almost perfect $W$ state. 

Furthermore, we showed that the present 
protocol works with the same efficiency whether or not
the number of qubits with Bob is equal to $M$. 
In this sense we can think of the present protocol as a way
to send and then transform an $M$-partite $W$ state with Alice to 
an $\tilde{M}$-partite $W$ state
with Bob, where $\tilde{M}\neq M$. This is achieved by properly setting the coupling constants of Alice's and Bob's qubits
with the end points of the spin chain such that the ratio of these couplings are
$\sqrt{M/\tilde{M}}$. We also showed how this protocol, consisting of 
$M+N+\tilde{M}$ qubits, can be mapped to an effective strictly linear chain
of $1+N+1$ qubits. 
The explicit map as well as the mathematical details justifying this map
were given in the main text.

We also studied the robustness of the present model to disorder and noise.
We studied time independent and dependent disorder as well as site (position)
dependent and independent disorder \cite{vie18,vie19}. For chains of the order
of a hundred qubits, we showed that the number of qubits (branches) with Alice and
Bob does not affect appreciably how the system responds to disorder. This is true as 
long as the number of branches is a fraction, say $10\%$ at most, of the size of 
the chain. In this scenario and working with the most severe type of disorder
(fluctuating disorder), we showed that fluctuations of the order of $1\%$ about the 
coupling constants of the ordered system do not affect considerably the transmission efficiency of the protocol. Also, for fluctuations going all the way up to $10\%$
about the optimal constants, we showed that it is very likely that we still have
genuine $W$-like multipartite entanglement reaching Bob, even though Bob's state is 
no longer close to the pure $W$ state sent by Alice. 
We have also studied in \ref{ap1} the case of spin chains with one thousand qubits,
where we showed that the greater the chain size the greater its sensitivity to
disorder.

Finally, we also introduced two types of noise in the present protocol, namely, the 
$\sigma_j^z\sigma_{j+1}^z$ interaction and transverse 
external magnetic fields acting on the spins. The noise operated in the same way
as disorder, being time and site dependent or independent. And similarly to 
the case of pure disorder, 
the worst case 
occurred for fluctuating noise. We showed that under certain circumstances 
the system's response to noise depends 
on the number of branches with Alice and Bob, a feature not seen when only disorder was 
present. In any case, we still had excellent state transmission efficiency for 
fluctuations of the order of $0.5\%$, when both disorder and the two types of noise 
affected simultaneously the system.

\section*{Acknowledgments}
GR thanks CNPq and CNPq/FAPERJ (State of Rio de Janeiro Research Foundation) for financial support through the National Institute of
Science and Technology for Quantum Information.

\appendix

\section{The matrix $\mathbf{F}$} 
\label{ap0}

The matrix $\mathbf{F}$, defining the system of linear equations (\ref{mateq}) that 
we need to solve to obtain the time evolution of
the single excitation system is
\begin{equation}
\mathbf{F} \hspace{-.1cm}=\hspace{-.1cm} -\frac{i2}{\hbar}\hspace{-.1cm}
\left(\hspace{-.2cm}
\begin{array}{ccccccccccccc}
0 \hspace{-.2cm} & \hspace{-.2cm} \cdots \hspace{-.2cm} & \hspace{-.2cm} 0 \hspace{-.2cm}  & \hspace{-.3cm}  J_{A_1} \hspace{-.3cm} & \hspace{-.2cm} 0 \hspace{-.2cm} & \hspace{-.2cm} 0 \hspace{-.2cm} & \hspace{-.2cm} 0 \hspace{-.2cm} & \hspace{-.2cm} 0 \hspace{-.2cm} & \hspace{-.2cm}\cdots \hspace{-.2cm} & \hspace{-.2cm} 0 \hspace{-.2cm} & \hspace{-.2cm} 0 \hspace{-.2cm} & \hspace{-.2cm} 0 \hspace{-.2cm} & \hspace{-.2cm} 0\\
\vdots \hspace{-.2cm} & \hspace{-.2cm} \ddots \hspace{-.2cm} & \hspace{-.2cm} \vdots & \hspace{-.2cm} \vdots \hspace{-.2cm} & \hspace{-.2cm} \vdots \hspace{-.2cm} & \hspace{-.2cm} \vdots \hspace{-.2cm} & \hspace{-.2cm} \vdots \hspace{-.2cm} & \hspace{-.2cm} \vdots \hspace{-.2cm} & \hspace{-.2cm}\ddots  \hspace{-.2cm} & \hspace{-.2cm} \vdots \hspace{-.2cm} & \hspace{-.2cm} \vdots \hspace{-.2cm} & \hspace{-.2cm} \vdots \hspace{-.2cm} & \hspace{-.2cm} \vdots\\
0 \hspace{-.2cm} & \hspace{-.2cm} \cdots \hspace{-.2cm} & \hspace{-.2cm} 0 \hspace{-.2cm} & \hspace{-.3cm} J_{A_{M}} \hspace{-.3cm} & \hspace{-.2cm} 0 \hspace{-.2cm} & \hspace{-.2cm} 0 \hspace{-.2cm} & \hspace{-.2cm} 0 \hspace{-.2cm} & \hspace{-.2cm} 0 \hspace{-.2cm} & \hspace{-.2cm} \cdots \hspace{-.2cm} & \hspace{-.2cm} 0 \hspace{-.2cm} & \hspace{-.2cm} 0 \hspace{-.2cm} & \hspace{-.2cm} 0 \hspace{-.2cm}  & \hspace{-.2cm} 0\\
J_{A_1} \hspace{-.3cm} & \hspace{-.2cm} \cdots \hspace{-.2cm} & \hspace{-.3cm} J_{A_{M}} \hspace{-.3cm}  & \hspace{-.2cm} 0 \hspace{-.2cm} & \hspace{-.2cm} J_{1,2} \hspace{-.2cm} & \hspace{-.2cm} 0 \hspace{-.2cm} & \hspace{-.2cm} 0 \hspace{-.2cm} & \hspace{-.2cm} 0 \hspace{-.2cm} & \hspace{-.2cm} \cdots \hspace{-.2cm} & \hspace{-.2cm} 0 \hspace{-.2cm} & \hspace{-.2cm} 0 \hspace{-.2cm} & \hspace{-.2cm} 0 \hspace{-.2cm}  & \hspace{-.2cm} 0\\
0 \hspace{-.2cm} & \hspace{-.2cm} \cdots \hspace{-.2cm} & \hspace{-.2cm} 0 \hspace{-.2cm} & \hspace{-.2cm} J_{1,2} \hspace{-.2cm} & \hspace{-.2cm} 0 \hspace{-.2cm} & \hspace{-.2cm} J_{2,3} \hspace{-.2cm} & \hspace{-.2cm} 0 \hspace{-.2cm} & \hspace{-.2cm} 0 \hspace{-.2cm} & \hspace{-.2cm} \cdots  \hspace{-.2cm} & \hspace{-.2cm} 0 \hspace{-.2cm} & \hspace{-.2cm} 0 \hspace{-.2cm} & \hspace{-.2cm} 0 \hspace{-.2cm} & \hspace{-.2cm} 0\\
0 \hspace{-.2cm} & \hspace{-.2cm} \cdots \hspace{-.2cm} & \hspace{-.2cm} 0 \hspace{-.2cm} & \hspace{-.2cm} 0 \hspace{-.2cm} & \hspace{-.2cm} J_{2,3} \hspace{-.2cm} & \hspace{-.2cm} 0 \hspace{-.2cm} & \hspace{-.2cm} J_{3,4} \hspace{-.2cm} & \hspace{-.2cm} 0 \hspace{-.2cm} & \hspace{-.2cm}\cdots  \hspace{-.2cm} & \hspace{-.2cm} 0 \hspace{-.2cm} & \hspace{-.2cm} 0 \hspace{-.2cm} & \hspace{-.2cm} 0 \hspace{-.2cm} & \hspace{-.2cm} 0\\
\vdots \hspace{-.2cm} & \hspace{-.2cm} \vdots \hspace{-.2cm} & \hspace{-.2cm} \vdots & \hspace{-.2cm} \vdots \hspace{-.2cm} & \hspace{-.2cm} \vdots \hspace{-.2cm} & \hspace{-.2cm} \vdots \hspace{-.2cm} & \hspace{-.2cm} \vdots \hspace{-.2cm} & \hspace{-.2cm} \vdots \hspace{-.2cm} & \hspace{-.2cm}\vdots  \hspace{-.2cm} & \hspace{-.2cm} \vdots \hspace{-.2cm} & \hspace{-.2cm} \vdots \hspace{-.2cm} & \hspace{-.2cm} \vdots \hspace{-.2cm} & \hspace{-.2cm} \vdots\\
0 \hspace{-.2cm} & \hspace{-.2cm} 0 \hspace{-.2cm} & \hspace{-.2cm} 0 & \hspace{-.2cm} 0 \hspace{-.2cm} & \hspace{-.2cm} \cdots \hspace{-.2cm}  & \hspace{-.2cm} 0 \hspace{-.2cm} & \hspace{-.4cm} J_{N-3,N-2} \hspace{-.4cm} & \hspace{-.2cm} 0 \hspace{-.2cm} & \hspace{-.4cm} J_{N-2,N-1} \hspace{-.4cm} & \hspace{-.2cm} 0 \hspace{-.2cm} & \hspace{-.2cm} 0 \hspace{-.2cm} & \hspace{-.2cm} \cdots \hspace{-.2cm} & \hspace{-.2cm} 0 \\
0 \hspace{-.2cm} & \hspace{-.2cm} 0 \hspace{-.2cm} & \hspace{-.2cm} 0 & \hspace{-.2cm} 0 \hspace{-.2cm} & \hspace{-.2cm} \cdots \hspace{-.2cm} & \hspace{-.2cm} 0 \hspace{-.2cm} & \hspace{-.2cm} 0 \hspace{-.2cm} & \hspace{-.3cm} J_{N-2,N-1} \hspace{-.4cm} & \hspace{-.2cm} 0 \hspace{-.2cm} & \hspace{-.4cm} J_{N-1,N} \hspace{-.4cm} & \hspace{-.2cm} 0 \hspace{-.2cm} & \hspace{-.2cm} \cdots \hspace{-.2cm} & \hspace{-.2cm} 0 \\
0 \hspace{-.2cm} & \hspace{-.2cm} 0 \hspace{-.2cm}& \hspace{-.2cm} 0  & \hspace{-.2cm} 0 \hspace{-.2cm} & \hspace{-.2cm} \cdots \hspace{-.2cm} & \hspace{-.2cm} 0 \hspace{-.2cm} & \hspace{-.2cm} 0 \hspace{-.2cm} & \hspace{-.2cm} 0 \hspace{-.2cm} & \hspace{-.4cm} J_{N-1,N} \hspace{-.4cm} & \hspace{-.2cm} 0 \hspace{-.2cm} & \hspace{-.4cm} J_{B_1} \hspace{-.4cm} & \hspace{-.2cm} \cdots \hspace{-.2cm} & \hspace{-.2cm} J_{B_{\tilde{M}}} \\
0 \hspace{-.2cm} & \hspace{-.2cm} 0 \hspace{-.2cm} & \hspace{-.2cm} 0 & \hspace{-.2cm} 0 \hspace{-.2cm} & \hspace{-.2cm} \cdots \hspace{-.2cm} & \hspace{-.2cm} 0 \hspace{-.2cm} & \hspace{-.2cm} 0 \hspace{-.2cm} & \hspace{-.2cm} 0 \hspace{-.2cm} & \hspace{-.2cm} 0 \hspace{-.2cm} & \hspace{-.4cm} J_{B_1} \hspace{-.4cm} & \hspace{-.2cm} 0 \hspace{-.2cm} & \hspace{-.2cm} \cdots \hspace{-.2cm} & \hspace{-.2cm} 0 \\
\vdots \hspace{-.2cm} & \hspace{-.2cm} \vdots \hspace{-.2cm} & \hspace{-.2cm} \vdots & \hspace{-.2cm} \vdots \hspace{-.2cm} & \hspace{-.2cm} \ddots \hspace{-.2cm} & \hspace{-.2cm} \vdots \hspace{-.2cm} & \hspace{-.2cm} \vdots \hspace{-.2cm} & \hspace{-.2cm} \vdots  \hspace{-.2cm} & \hspace{-.2cm} \vdots \hspace{-.2cm} & \hspace{-.2cm} \vdots \hspace{-.2cm} & \hspace{-.2cm} \vdots \hspace{-.2cm} & \hspace{-.2cm} \ddots \hspace{-.2cm} & \hspace{-.2cm} \vdots\\
0 \hspace{-.2cm} & \hspace{-.2cm} 0 \hspace{-.2cm} & \hspace{-.2cm} 0 & \hspace{-.2cm} 0 \hspace{-.2cm} & \hspace{-.2cm} \cdots \hspace{-.2cm} & \hspace{-.2cm} 0 \hspace{-.2cm} & \hspace{-.2cm} 0 \hspace{-.2cm} & \hspace{-.2cm} 0 \hspace{-.2cm} & \hspace{-.2cm} 0 \hspace{-.2cm} & \hspace{-.4cm} J_{B_{\tilde{M}}} \hspace{-.4cm} & \hspace{-.2cm} 0 \hspace{-.2cm} & \hspace{-.2cm} \cdots \hspace{-.2cm} & \hspace{-.2cm} 0
\end{array}
\hspace{-.2cm}\right).
\label{matriz}
\end{equation}

\section{Proof of Eqs.~(\ref{fide-fido0}) and (\ref{fide-fido1})} 
\label{apA}

Our goal here is to obtain an analytic formula of the fidelity $F(t)$, Eq.~(\ref{fid1}),
for the effective $N+2$ strictly linear chain when $J_m \gg J$ (asymptotic regime).  
We also assume the system satisfies the conditions given in 
Eqs.~(\ref{cond1}) and (\ref{cond2}). To obtain the asymptotic formula, we 
explicitly diagonalize the Hamiltonian assuming $J_m \gg J$. 
Then, by working in the basis that diagonalizes the Hamiltonian, 
the fidelity in the 
asymptotic regime can be computed in a closed formula. 

Let us start writing the Hamiltonian (\ref{ham0}) in the diagonal basis,
\begin{equation}
H = \sum_{k=1}^{N+2} E_k |E_k \rangle \langle E_k|,
\end{equation}
where $|E_k \rangle$ is the $k$-th eigenstate of $H$ with energy $E_k$.\footnote{\label{foot3}
Working out analytically the cases for small values of $N+2$ or numerically
solving for values of $N+2$ up to hundreds of qubits, we see the following pattern: 
(1) there is no degeneracy in the system; (2) for even $N+2$ we have $(N+2)/2$ pairs of 
eigenvalues $(-E_k,E_k)$; (3) for odd $N+2$ we have a central null eigenvalue and 
$(N+1)/2$ pairs $(-E_k,E_k)$; and (4) this trend is true whether or not $J_m=J$.}  
Using the notation given by Eq.~(\ref{l2}), 
the fidelity (\ref{fid1}) becomes
\begin{equation}
F(t) = |\langle 1_{N+2}|e^{-iHt/\hbar}|1_1\rangle|^2 
= \left|\sum_{k=1}^{N+2}e^{-iE_kt/\hbar}\langle 1_{N+2}|E_k\rangle\langle E_k|1_1\rangle\right|^2,
\label{fid}
\end{equation}
where the last expression above comes from inserting the identity operator 
$\sum_{k=1}^{N+2} |E_k \rangle \langle E_k| =\mathbb{1}$.

To obtain the eigenvalues $E_k$ we must solve the following characteristic equation,
\begin{equation}
D = \det (H - E_k\mathbb{1}) = 0,
\label{det}
\end{equation}
where ``$\det$'' stands for the determinant, $\mathbb{1}$ is the identity matrix,
and $H$ is the Hamiltonian of the effective 
strictly linear chain (Eq.~(\ref{hamN0}) with $J_{A,1}=J_{N,B}=J$ and $J_{j,j+1}=J_m$).

Working in the basis $\{|1_1\rangle, \ldots, |1_{N+2}\rangle\}$, we can use
the Laplace expansion (cofactor expansion) to write Eq.~(\ref{det}) as
\begin{equation}
D= 16(J^2_m \cos^2 \theta_k D_N - 2 J^2 J_m  \cos \theta_k  D_{N-1} 
+ J^4 D_{N-2})=0, \label{det1a}
\end{equation}
where
\begin{eqnarray}
E_k&=&-4 J_m \cos \theta_k, \label{ek}\\
D_N&=& \det (H_N - E_k\mathbb{1}).
\end{eqnarray}
Here $H_N$ represents the strictly linear XX model composed of $N$ qubits 
(cf. Eq.~(\ref{hamNN})). 

Applying the Laplace expansion to $D_N$ we obtain a recursive relation similar to
Eq.~(\ref{det1a}). Solving it we get 
\begin{equation}
D_N = (2J_m)^N  \sin[(N+1)\theta_k]/\sin\theta_k. 
\label{dn}
\end{equation}
Inserting Eq.~(\ref{dn}) into (\ref{det1a}) the characteristic equation becomes, up
to an overall irrelevant factor,
\begin{equation}
4 \cos^2 \theta_k \sin[(N+1)\theta_k] -4 \left(\frac{J}{J_m}\right)^2 
\cos \theta_k \sin[N \theta_k] + \left(\frac{J}{J_m}\right)^4 \sin[(N-1) \theta_k]=0. \label{det2}
\end{equation}
For arbitrary values of $J$ and $J_m$ we could not solve analytically 
Eq.~(\ref{det2}). However, for $J_m >> J$ we can neglect its last two terms obtaining
\begin{equation}
\cos^2{\theta_k}\sin[(N+1)\theta_k] = 0, 
\label{zero}
\end{equation}
whose solutions are $\theta_1=\theta_{N+2}=\pi/2$  
and $\theta_k=(k-1)\pi/(N+1)$, $k=2,\ldots, N+1$.
This leads to the following approximation to the eigenenergies of the system,
\begin{eqnarray}
E_1&=&E_{N+2} \approx 0, \label{eofid}\\
E_k &\approx& -4J_m \cos \left(\frac{(k-1) \pi}{N+1}\right), \hspace{.25cm} \mbox{where} 
\hspace{.25cm} k = 2,\ldots, N+1. \label{ekfid}
\end{eqnarray}

Note that $E_k$, $k = 2,\ldots, N+1$, are the eigenvalues of the Hamiltonian for a 
strictly linear chain of $N$ spins described by the XX model. 
It is the Hamiltonian we get by setting $J=0$ at the effective
model, i.e., qubits $1$ and $N+2$ do not interact with the end points of the chain.
Those values for $E_k$ are consistent with the level of approximation we are interested in.
However, to obtain non-null values for $E_1$ and $E_{N+2}$, we need to solve Eq.~(\ref{det2}) keeping the next non-zero relevant term. 

First, we should note that the values of $\theta_1$ and $\theta_{N+2}$ are very close
to $\pi/2$ since they come from solving for $\cos^2\theta_k=0$ in Eq.~(\ref{zero}).
Since we must have $E_1=-E_{N+2}$ (see discussion in footnote \ref{foot3}), 
$\theta_1 \approx \pi/2 - x$ and $\theta_{N+2} \approx \pi/2 +x$ if we choose $0<x\ll 0$,
$J_m>0$, and $E_{N+2}$ $=-4J_m$ $\cos\theta_{N+2}>0$. For definiteness, 
we will show how to get $\theta_1$ and hence $E_1$.

For $N$ odd we have to solve Eq.~(\ref{det2}) keeping the first two terms, 
\begin{equation}
\cos \theta_1 \sin[(N+1)\theta_1] - \left(\frac{J}{J_m}\right)^2 
\sin[N \theta_1] = 0.
\label{eqOdd}
\end{equation}
Writing $\theta_1 = \pi/2 - x$ and Taylor expanding to first order in $x$ we have
$\cos(\theta_1)\approx x$, $\sin[(N+1)\theta_1] = (N+1)x\sin[N\pi/2]$, and
$\sin(N\theta_1)\approx \sin(N\pi/2)$. Inserting these Taylor expansions into Eq.~(\ref{eqOdd}) we get
\begin{equation}
x = \frac{1}{\sqrt{N+1}}\frac{J}{J_m} \longrightarrow 
\theta_1 = \frac{\pi}{2} - \frac{1}{\sqrt{N+1}}\frac{J}{J_m}.
\label{xOdd}
\end{equation}
To get $E_1$ for odd $N$ we insert Eq.~(\ref{xOdd}) into (\ref{ek}) and Taylor expand
the cosine to first order in $J/J_m$. This leads to
\begin{equation}
E_1 \approx  -\frac{4J}{\sqrt{N+1}}, \hspace{.25cm}\mbox{for N odd}. \label{det6}
\end{equation}

For $N$ even we have $\sin[(N\pm 1)\theta_1]
= \pm \cos(N\pi/2) + \mathcal{O}(x^2)$ and 
$\sin[N\theta_1] = -Nx\cos(N\pi/2) + \mathcal{O}(x^3)$
since $\sin[N\pi/2] = 0$. Thus, if we use only the first two terms of 
Eq.~(\ref{det2}) we get $x=0$ to order $x$. We must go to second order in $x$ and also 
keep the third term of Eq.~(\ref{det2}) to make progress. 
Using the previous Taylor expansions and noting that $\cos(\theta_1)\approx x+\mathcal{O}(x^3)$, 
Eq.~(\ref{det2}) becomes to leading order in $x$,
\begin{equation}
x^2 \left((N-1)^2+8N J_m^2/J^2+8
J_m^4/J^4\right)-2=0. \label{detEven}
\end{equation}
Solving for $x$ and Taylor expanding the solution in powers of $J/J_m$ we get to leading order
\begin{equation}
x = \frac{1}{2}\left(\frac{J}{J_m}\right)^2 \longrightarrow \theta_1 = \frac{\pi}{2} - \frac{1}{2}\left(\frac{J}{J_m}\right)^2.
\label{xEven}
\end{equation}
Inserting Eq.~(\ref{xEven}) into (\ref{ek}) and Taylor expanding we get 
\begin{equation}
E_1 \approx  -\frac{2J^2}{J_m}, \hspace{.25cm}\mbox{for N even}. \label{det4}
\end{equation}

We now turn to the computation of the eigenvectors. In the basis 
$\{|1_1\rangle,\ldots |1_{N+2}\rangle\}$ the Hamiltonian for the effective 
linear chain can be written as
\begin{equation}
H = H_A + H_N + H_B,
\label{hamE}
\end{equation}
where
\begin{eqnarray}
H_A &=& 2J|1_1\rangle\langle 1_2| + h.c., \\
H_N &=& 2J_m\sum_{j=2}^{N}(|1_j\rangle\langle 1_{j+1}| + h.c.), \\
H_B &=& 2J|1_{N+1}\rangle\langle 1_{N+2}| + h.c., 
\end{eqnarray}
with $h.c.$ denoting the Hermitian conjugate of the term preceding it. 
We also write the unnormalized eigenvector of $H$ as
\begin{equation}
|\tilde{E}_k\rangle = \sum_{j=1}^{N+2}a_{k,j}|1_j\rangle.
\label{etilde}
\end{equation}
To obtain the coefficients of Eq.~(\ref{etilde}) we have to solve
\begin{equation}
H |\tilde{E}_k\rangle = E_k |\tilde{E}_k\rangle.
\label{eigenequation}
\end{equation}
Using Eqs.~(\ref{hamE}) and (\ref{etilde}),
the left hand side of Eq.~(\ref{eigenequation}) becomes 
after a little algebra,
\begin{eqnarray}
 H |\tilde{E}_k\rangle &=& \sum_{j=3}^{N}[2J_m(a_{k,j+1}+a_{k,j-1})]|1_j\rangle
 + 2Ja_{k,2}|1_1\rangle + 2Ja_{k,N+1}|1_{N+2}\rangle \nonumber \\
 & & + [2J_ma_{k,3}+2Ja_{k,1}]|1_2\rangle + [2J_ma_{k,N}+2Ja_{k,N+2}]|1_{N+1}\rangle,
\label{lhs}
 \end{eqnarray}
while the right hand side can be written as
\begin{eqnarray}
 E_k |\tilde{E}_k\rangle &=& \sum_{j=3}^{N}[E_ka_{k,j}]|1_j\rangle
 + E_ka_{k,1}|1_1\rangle + E_ka_{k,N+2}|1_{N+2}\rangle \nonumber \\
 & & + E_ka_{k,2}|1_2\rangle + E_ka_{k,N+1}|1_{N+1}\rangle.
 \label{rhs}
\end{eqnarray}
Comparing Eqs.~(\ref{lhs}) and (\ref{rhs}) we obtain the following set of equations
whose solution will give $a_{k,j}$,
\begin{eqnarray}
2J_m(a_{k,j+1}+a_{k,j-1}) &=& E_ka_{k,j}, \hspace{.25cm} \mbox{for} \hspace{.25cm}
3\leq j \leq N, \label{a26}\\
2Ja_{k,2} & = & E_ka_{k,1}, \label{a27}\\
2Ja_{k,N+1} & = & E_ka_{k,N+2}, \label{a28}\\
2J_ma_{k,3}+2Ja_{k,1} & = & E_ka_{k,2},\label{a29}\\
2J_ma_{k,N}+2Ja_{k,N+2} & = & E_ka_{k,N+1}. \label{a30}
\end{eqnarray}

%

If we write the normalized eigenvector of $H$ as
\begin{equation}
|E_k\rangle = A_k\sum_{j=1}^{N+2}a_{k,j}|1_j\rangle
\label{e}
\end{equation}
and insert it into Eq.~(\ref{fid}), we get for the fidelity
\begin{equation}
F(t) = 
\left|
\sum_{k=1}^{N+2}e^{-iE_kt/\hbar}A_k^2 a_{k,1}^* a_{k,N+2}
\right|^2,
\label{fid-ap1}
\end{equation}
where $A_k$ is the normalization constant for the state $|E_k\rangle$. 
Looking at 
Eq.~(\ref{fid-ap1}) we realize that only the coefficients $a_{k,1}$ and $a_{k,N+2}$
are the relevant ones in the computation of the fidelity. Moreover, 
the eigenvectors' coefficients $a_{k,j}$ 
are computed to leading order by noting that there are two classes of eigenvalues. 
For $J_m\gg J$ we either have   
$E_k\approx \mathcal{O}(J_m)$ or  $E_k\approx 0 + \mathcal{O}(J)$
(cf. Eqs.~(\ref{eofid}) and (\ref{ekfid})).

When $E_k\approx \mathcal{O}(J_m)$, Eqs.~(\ref{a27}) and (\ref{a28}) give 
\begin{equation}
a_{k,1} =  \mathcal{O}\left(\frac{J}{J_m}a_{k,2}\right) \hspace {.25cm} \mbox{and} 
\hspace{.25cm} a_{k,N+2} =  \mathcal{O}\left(\frac{J}{J_m}a_{k,N+1}\right)
\label{a27e28}
\end{equation}
while Eqs.~(\ref{a29}) and (\ref{a30}) together with 
Eq.~(\ref{a27e28}) lead to 
\begin{equation}
a_{k,2} \approx  a_{k,3} \hspace {.25cm} \mbox{and} 
\hspace{.25cm} a_{k,N} \approx a_{k,N+1}.
\label{a29e30}
\end{equation}
If we now use Eq.~(\ref{a26}) and Eqs.~(\ref{a27e28}) and (\ref{a29e30}), 
we obtain that the remaining $a_{k,j}$, for $j=2,\ldots, N+1$, almost always 
satisfy 
\begin{equation}
a_{k,j} \approx \mathcal{O}\left(a_{k,2}\right) \hspace {.25cm} \mbox{or} 
\hspace{.25cm}
a_{k,j} \approx \mathcal{O}\left(a_{k,N+1}\right).
\label{a35}
\end{equation}
Thus, looking at Eqs.~(\ref{a27e28})-(\ref{a35}), we get that in the asymptotic limit 
($J/J_m\gg 0$) the relative weights of the coefficients $a_{k,1}$ and $a_{k,N+2}$ in 
the expansion of the eigenvector $|E_k\rangle$ are 
negligible when compared to most of the other coefficients. This implies that in the 
formula for the fidelity, Eq.~(\ref{fid-ap1}), the contribution coming from the eigenvectors
with eigenenergies of the order of $J_m$ do not contribute much.

Indeed, when 
$E_k\approx 0 + \mathcal{O}(J)$ we can repeat the previous analysis using 
Eqs.~(\ref{a26})-(\ref{a30}) to obtain
\begin{equation}
a_{k,1} =  \mathcal{O}\left(\frac{J_m}{J}a_{k,3}\right) \hspace {.25cm} \mbox{and} 
\hspace{.25cm} a_{k,N+2} =  \mathcal{O}\left(\frac{J_m}{J}a_{k,N}\right),
\label{a27e28b}
\end{equation}
while the remaining coefficients are either of order of $a_{k,1}$ and $a_{k,N+2}$
or of order $(J/J_m)a_{k,1}$ and $(J/J_m)a_{k,N+2}$. Therefore, in the asymptotic limit
the contributions of $a_{k,1}$ and $a_{k,N+2}$ are relevant to the calculation of the
fidelity and dominate the ones coming
from the case where $E_k\approx \mathcal{O}(J_m)$ by at least one order of magnitude in
$J/J_m$.

For $N$ even we only have two cases in which $E_k\approx 0 + \mathcal{O}(J)$, namely,
$E_1$ and $E_{N+2}$. Using Eq.~(\ref{det4}) for $E_1$ and solving Eqs.~(\ref{a26})-(\ref{a30}), we obtain to leading order
\begin{eqnarray}
a_{1,1} &=& a_{1,N+2}, \label{a37} \\
a_{1,j_{even}} &=& (J/J_m)\cos(j_{even}\pi/2) a_{1,1}, 
\hspace {.25cm} \mbox{for} \hspace{.25cm} 2\leq j_{even} \leq N+1, \label{a38}\\
a_{1,j_{odd}} &=& (J/J_m)\sin(j_{odd}\pi/2) a_{1,1}, 
\hspace {.25cm} \mbox{for} \hspace{.25cm} 2\leq j_{odd} \leq N+1.
\label{a39}
\end{eqnarray}
Noting that $E_{N+2}=-E_1$ we similarly get
\begin{eqnarray}
a_{N+2,1} &=& -a_{N+2,N+2}, \label{a40} \\
a_{N+2,j_{even}} &=& -(J/J_m)\cos(j_{even}\pi/2) a_{N+2,1}, 
\hspace {.25cm} \mbox{for} \hspace{.25cm} 2\leq j_{even} \leq N+1, \label{a41}\\
a_{N+2,j_{odd}} &=& (J/J_m)\sin(j_{odd}\pi/2) a_{N+2,1}, 
\hspace {.25cm} \mbox{for} \hspace{.25cm} 2\leq j_{odd} \leq N+1.
\label{a42}
\end{eqnarray}
Equations (\ref{a37})-(\ref{a42}) show that only $a_{1,1}$, $a_{1,N+2}$,
$a_{N+2,1}$, and $a_{N+2,N+2}$ survive when $J/J_m\rightarrow 0$. To leading order we 
thus have
\begin{equation}
|E_1\rangle \approx (1/\sqrt{2})(|1_1\rangle + |1_{N+2}\rangle)
\hspace {.25cm} \mbox{and} \hspace{.25cm}
|E_{N+2}\rangle \approx (1/\sqrt{2})(|1_1\rangle - |1_{N+2}\rangle).
\end{equation}
Comparing Eq.~(\ref{e}) with $|E_1\rangle$ and $|E_{N+2}\rangle$ as given above, we can 
compute Eq.~(\ref{fid-ap1}). After a little algebra  we get
\begin{equation}
F(t) =
\sin^2\left[2\left(\frac{J}{J_m}\right) \frac{Jt}{\hbar}\right], \hspace{.25cm}\mbox{for N even},
\label{fid-ap-even2}
\end{equation}
which is Eq.~(\ref{fide-fido0}) we wanted to prove.

For $N$ odd we also have $E_k\approx 0 +\mathcal{O}(J)$ for $k=1$ and $k=N+2$ and, 
in addition, $E_{(N+3)/2}=0$, the central eigenvalue which can be approximated as being
exactly zero 
(see Eq.~(\ref{ekfid}) and the discussion in footnote \ref{foot3}). Using Eq.~(\ref{det6})
for $E_1$ we can solve Eqs.~(\ref{a26})-(\ref{a30}). To leading order we have
\begin{eqnarray}
a_{1,1} &=& - a_{1,N+2}, \label{a37o} \\
a_{1,j_{even}} &=& (2/\sqrt{N+1})\cos(j_{even}\pi/2) a_{1,1}, 
\hspace {.25cm} \mbox{for} \hspace{.25cm} 2\leq j_{even} \leq N+1, \label{a38o}\\
a_{1,j_{odd}} &=& (J/J_m)[1-2(j-1)/(N+1)]\sin(j_{odd}\pi/2) a_{1,1}, 
\hspace {.25cm} \mbox{for} \hspace{.25cm} 2\leq j_{odd} \leq N+1.
\label{a39o}
\end{eqnarray}
Similarly for $E_{N+2}=-E_1$ we get
\begin{eqnarray}
a_{N+2,1} &=& -a_{N+2,N+2}, \label{a40o} \\
a_{N+2,j_{even}} &=& -(2/\sqrt{N+1})\cos(j_{even}\pi/2) a_{N+2,1}, 
\hspace {.25cm} \mbox{for} \hspace{.25cm} 2\leq j_{even} \leq N+1, \label{a41o}\\
a_{N+2,j_{odd}} &=& (J/J_m)[1-2(j-1)/(N+1)]\sin(j_{odd}\pi/2), 
\hspace {.25cm} \mbox{for} \hspace{.25cm} 2\leq j_{odd} \leq N+1.
\label{a42o}
\end{eqnarray}
If we now set $E_{(N+3)/2}=0$ in Eqs.~(\ref{a26})-(\ref{a30}) we obtain
\begin{eqnarray}
a_{(N+3)/2,1} &=& a_{(N+3)/2,N+2}, \label{a37c} \\
a_{(N+3)/2,j_{even}} &=& 0, 
\hspace {.25cm} \mbox{for} \hspace{.25cm} 2\leq j_{even} \leq N+1, \label{a38c}\\
a_{(N+3)/2,j_{odd}} &=& (J/J_m)\sin(j_{odd}\pi/2) a_{(N+3)/2,1}, 
\hspace {.25cm} \mbox{for} \hspace{.25cm} 2\leq j_{odd} \leq N+1.
\label{a39c}
\end{eqnarray}

Using Eqs.~(\ref{a37o})-(\ref{a39c}) we can compute the three corresponding eigenvectors.
In the limit where $J/J_m\ll 1$, the normalized eigenvectors are to leading order
\begin{eqnarray}
|E_1\rangle &=& \frac{1}{2}\left[ |1_1\rangle + 
\sum_{j_{even}=2}^{N+1}\frac{2}{\sqrt{N+1}}
\cos\left(j_{even}\frac{\pi}{2}\right)|1_j\rangle 
- |1_{N+2}\rangle\right], \label{e1o} \\
|E_{N+2}\rangle &=& \frac{1}{2}\left[ |1_1\rangle - 
\sum_{j_{even}=2}^{N+1}\frac{2}{\sqrt{N+1}}
\cos\left(j_{even}\frac{\pi}{2}\right)|1_j\rangle 
- |1_{N+2}\rangle\right], \label{eN2o} \\
|E_{(N+3)/2}\rangle &=& \frac{1}{\sqrt{2}}\left( |1_1\rangle 
+|1_{N+2}\rangle\right). \label{eN3o} 
\end{eqnarray}

Finally, with Eqs.~(\ref{e}), (\ref{fid-ap1}), and (\ref{e1o})-(\ref{eN3o}) we get for the
fidelity 
\begin{equation}
F(t) =
\sin^4\left( \frac{2}{\sqrt{N+1}}\frac{Jt}{\hbar}\right), \hspace{.25cm}\mbox{for N odd},
\label{fid-ap-odd2}
\end{equation}
proving thus Eq.~(\ref{fide-fido1}).

\section{Fluctuating disorder with an $N = 1000$ spin chain} \label{ap1}

The computational resources to deal with $N=1000$ qubits are more demanding and, therefore,
we implement only $100$ realizations of disorder for each value of $p$ instead of the $1000$ ones 
for the $N=100$ qubit chain.
We also limit our analyses to the worst possible scenario, namely, fluctuating disorder and,
as before, we set $\tau=10\% t_{\!_{MAX}}$.

\begin{figure}[!ht] \begin{center}
\includegraphics[width=8cm]{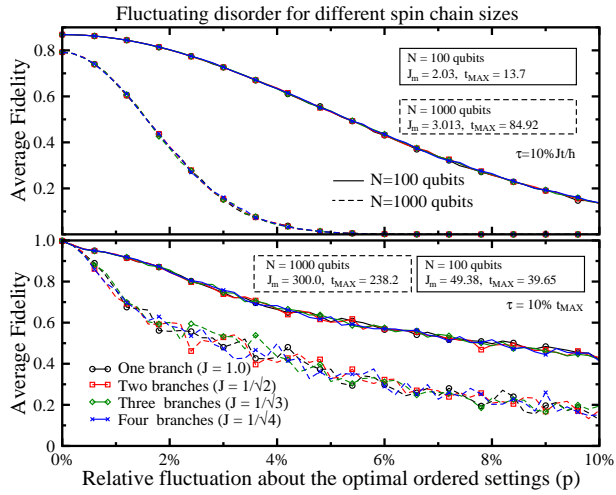}
\caption{\label{fig7}
Comparison between the $N = 100$ qubit chain (continuous lines) with the $N = 1000$ qubit chain (dashed lines). The meaning of all quantities is equal to the ones shown
in the figures of the main text. The values of $t_{\!_{MAX}}$, the optimal time
yielding the greatest transmission fidelity for the ordered model, are
given in the figure for each case studied. The optimal values for $J_m$ for the 
$N=100$ case are the same as in the main text while for $N=1000$ we show the
optimal $J_m$ when $J_m\leq 5.0$ (upper panel) and $J_m\leq 300.0$ (lower panel).}
\end{center} 
\end{figure}

As can be seen in Fig.~\ref{fig7}, the more qubits we have connecting Alice's and Bob's branches, the more intensely the transmission efficiency is affected by the presence of disorder. 
Indeed, as we add more qubits to the quantum wire connecting Alice and Bob 
we increase the number of instances in which disorder and noise can act, reducing thus the transmission capacity 
of the protocol.

\begin{thebibliography}{200}

\bibitem{ben00} C. H. Bennet, D. P. DiVincenzo, Nature (London) 404 (2000) 247.

\bibitem{bos03} S. Bose, Phys. Rev. Lett. 91 (2003) 207901.

\bibitem{chr04} M. Christandl, N. Datta, A. Ekert, A. J. Landahl, Phys. Rev. Lett. 92 (2004) 187902.

\bibitem{alb04} C. Albanese, M. Christandl, N. Datta, A. Ekert, Phys. Rev. Lett. 93 (2004) 230502.

\bibitem{chr05} M. Christandl, N. Datta, T. C. Dorlas, A. Ekert, A. Kay, A. J. Landahl, Phys. Rev. A 71 (2005) 032312.

\bibitem{nik04} G. M. Nikolopoulos, D. Petrosyan, P. L. Lambropoulos, J. Phys.: Condens. Matter 16 (2004) 4991.

\bibitem{sub04} V. Subrahmanyam, Phys. Rev. A 69 (2004) 034304.

\bibitem{woj05} A. W\'ojcik, T. \L{}uczak, P. Kurzy\'nski, A. Grudka, T. Gdala, M. Bednarska, Phys. Rev. A 72 (2005) 034303.

\bibitem{kar05} P. Karbach, J. Stolze, Phys. Rev. A 72 (2005) 030301.

\bibitem{shi05} T. Shi, Y. Li, Z. Song, Ch.-P. Sun, Phys. Rev. A 71 (2005) 032309.

\bibitem{har06} M. J. Hartmann, M. E. Reuter, M. B. Plenio, New J. Phys. 8 (2006) 94.

\bibitem{huo08} M. X. Huo, Y. Li, Z. Song, C. P. Sun, Europhys. Lett. 84 (2008) 30004.

\bibitem{gua08} G. Gualdi, V. Kostak, I. Marzoli, P. Tombesi, Phys. Rev. A 78 (2008) 022325.

\bibitem{ban10} L. Banchi, T. J. G. Apollaro, A. Cuccoli, R. Vaia, P. Verrucchi, Phys. Rev. A 82 (2010) 052321.

\bibitem{ban11} L. Banchi, T. J. G. Apollaro, A. Cuccoli, R. Vaia, P. Verrucchi, New J. Phys. 13 (2011) 123006.

\bibitem{kur11} P. Kurzy\'nski, A. W\'ojcik, Phys. Rev. A 83 (2011) 062315.

\bibitem{god12} C. Godsil, S. Kirkland, S. Severini, J. Smith, Phys. Rev. Lett. 109 (2012) 050502.

\bibitem{apo12} T. J. G. Apollaro, L. Banchi, A. Cuccoli, R. Vaia, P. Verrucchi, Phys. Rev. A 85 (2012) 052319.

\bibitem{lor13} S. Lorenzo, T. J. G. Apollaro, A. Sindona, F. Plastina, Phys. Rev. A 87 (2013) 042313.

\bibitem{apo13} T. J. G. Apollaro, S. Lorenzo, F. Plastina, Int. J. Mod. Phys. B 27 (2013) 1345035. 

\bibitem{kor14} K. Korzekwa, P. Machnikowski, P. Horodecki, Phys. Rev. A 89 (2014) 062301.

\bibitem{shi15} Z. C. Shi, X. L. Zhao, X. X. Yi, Phys. Rev. A 91 (2015) 032301.

\bibitem{pou15} S. Pouyandeh, F. Shahbazi, Int. J. Quantum Inform. 13 (2015) 1550030.

\bibitem{zha16} X.-P. Zhang, B. Shao, S. Hu, J. Zou, L.-A. Wu, Ann. Phys. (NY) 375 (2016) 435.

\bibitem{che16} X. Chen, R. Mereau, D. L. Feder, Phys. Rev. A 93 (2016) 012343.

\bibitem{lof16} N. J. S. Loft, O. V. Marchukov, D. Petrosyan, N. T. Zinner, New J. Phys. 18(4) (2016) 04511.
 
\bibitem{est17} M. P. Estarellas, I. D'Amico, T. P. Spiller, Sci. Rep. 7 (2017) 42904.
 
\bibitem{est17b} M. P. Estarellas, I. D'Amico, T. P. Spiller, Phys. Rev. A 95 (2017) 042335.
 
\bibitem{li05} Y. Li, T. Shi, B. Chen, Z. Song, C.-P. Sun, Phys. Rev. A 71 (2005) 022301.

\bibitem{apo19} T. J. G. Apollaro, G. M. A. Almeida, S. Lorenzo, A. Ferraro, S. Paganelli, Phys. Rev. A 100 (2019) 052308.

\bibitem{apo15} T. J. G. Apollaro, S. Lorenzo, A. Sindona, S. Paganelli, G. L. Giorgi, F. Plastina, Physica Scripta T165 (2015) 014036.

\bibitem{lor16} S. Lorenzo, T. J. G. Apollaro, A. Trombettoni, S. Paganelli, Available from: arXiv:1610.03248.

\bibitem{cir97} J. I. Cirac, P. Zoller, H. Kimble, H. Mabuchi, Phys. Rev. Lett. 78 (1997) 3221.

\bibitem{ple04} M. B. Plenio, J. Hartley, J. Eisert, New J. Phys. 6 (2004) 36.

\bibitem{sem05} M. B. Plenio, F. L. Semi\~ao, New J. Phys. 7 (2005) 73.

\bibitem{nic16} F. Nicacio, F. L. Semi\~ao, Phys. Rev. A 94 (2016) 012327.

\bibitem{vie18} R. Vieira, G. Rigolin, Phys. Lett. A 382 (2018) 2586.

\bibitem{vie19} R. Vieira, G. Rigolin, Quantum Inf. Process. 18 (2019) 135.

\bibitem{che19} W. J. Chetcuti, C. Sanavio, S. Lorenzo, T. J. G. Apollaro,  
New J. Phys. 22 (2020) 033030.

\bibitem{apo19b} T. J. G. Apollaro, C. Sanavio, W. J. Chetcuti, S. Lorenzo, 
Phys. Lett. A 384 (2020) 126306.

\bibitem{mar16} O. V. Marchukov, A. G. Volosniev, M. Valiente, D. Petrosyan, N. T. Zinner, Nature Commun. 7 (2016) 13070.

\bibitem{dur00} W. D\"{u}r, G. Vidal, J. I. Cirac, Phys. Rev. A 62 (2000) 062314.

\bibitem{joo02} J. Joo, J. Lee, J. Jang, Y. J. Park, Available from: arXiv:quant-ph/0204003.

\bibitem{wan07} J. Wang, Q. Zhang, C. J. Tang, ZHANG Quan, Commun. Theor. Phys. 48 (2007) 637.

\bibitem{gor03} V.N. Gorbachev, A.I. Trubilko, A.A. Rodichkina, Phys. Lett. A 314 (2003) 267.

\bibitem{kay05} A. Kay, M. Ericsson, New J. Phys. 7 (2005) 143

\bibitem{pem11} P. J. Pemberton-Ross, A. Kay, Phys. Rev. Lett. 106 (2011) 020503.

\bibitem{lor15} S. Lorenzo, T. J. G. Apollaro, S. Paganelli, G. M. Palma, F. Plastina, Phys. Rev. A 91 (2015) 042321.

\bibitem{vie13} R. Vieira, E. P. M. Amorim, G. Rigolin, Phys. Rev. Lett. 111 (2013) 180503.

\bibitem{vie14} R. Vieira, E. P. M. Amorim, G. Rigolin, Phys. Rev. A 382 (2014) 2586.

\bibitem{chi05} G. De Chiara, D. Rossini, S. Montangero, R. Fazio, Phys. Rev. A 72 (2005) 012323.

\bibitem{fit05} J. Fitzsimons, J. Twamley, Phys. Rev. A 72 (2005) 050301.

\bibitem{bur05} D. Burgarth, S. Bose, New J. Phys. 7 (2005) 135.

\bibitem{pet10} D. Petrosyan, G. M. Nikolopoulos, P. Lambropoulos, Phys. Rev. A 81 (2010) 042307.

\bibitem{zwi11} A. Zwick, G. A. \'Alvarez, J. Stolze, O. Osenda, Phys. Rev. A 84 (2011) 022311; 85 (2012) 012318.

\bibitem{bru12} M. Bruderer, K. Franke, S. Ragg, W. Belzig, D. Obreschkow, Phys. Rev. A 85 (2012) 022312.

\bibitem{nik13} G. M. Nikolopoulos, Phys. Rev. A 87 (2013) 042311.

\bibitem{kur14} A. Zwick, G. A. \'Alvarez, G. Bensky, G. Kurizki, New. J. Phys. 16 (2014) 065021.

\bibitem{ash15} S. Ashhab, Phys. Rev. A 92 (2015) 062305.

\bibitem{pav16} A. K. Pavlis, G. M. Nikolopoulos, P. Lambropoulos, Quantum Inf. Process. 15 (2016) 2553.

\bibitem{ben03} S. C. Benjamin, S. Bose, Phys. Rev. Lett. 90 (2003) 247901.

\bibitem{ron16} R. Ronke, M. P. Estarellas, I. D'Amico, T. P. Spiller, T. Miyadera, Eur. Phys. J. D 70 (2016) 189.

\bibitem{lyr17} G. M. A. Almeida, F. A. B. F. de Moura, T. J. G. Apollaro, M. L. Lyra, Phys. Rev. A 96 (2017) 032315.

\bibitem{lyr17a} G. M. A. Almeida, F. A. B. F. de Moura, M. L. Lyra, Phys. Lett. A 382
(2018) 1335.

\bibitem{lyr17b} G. M. A. Almeida, F. A. B. F. de Moura, M. L. Lyra, Quantum Inf. Process. 18 
(2019) 41.

\bibitem{kie07} N. Kiesel, C. Schmid, G. T\'oth, E. Solano, H. Weinfurter, Phys. Rev. Lett. 98 (2007) 063604.

\bibitem{woo98} W. K. Wootters, Phys. Rev. Lett. 80 (1998) 2245.

\end{thebibliography}
\end{document}